\documentclass[a4paper,fleqn,usenatbib]{mnras}

\usepackage{newtxtext,newtxmath}
\usepackage[T1]{fontenc}
\usepackage{ae,aecompl}
\usepackage{gensymb}

\usepackage{graphicx}	% Including figure files
\usepackage{amsmath}	% Advanced maths commands
\usepackage{amssymb}	% Extra maths symbols
\usepackage{breakurl} 
\usepackage{textcomp}
\usepackage{subfig}
\usepackage{natbib}
\usepackage{fontenc}
\usepackage{epsf}
\usepackage{epstopdf}
\usepackage{caption}
\usepackage{color}
\usepackage{transparent}
\usepackage[flushleft]{threeparttable} % http://ctan.org/pkg/threeparttable
\usepackage{booktabs,caption}
\title[Small electron acceleration episodes]{Small electron acceleration episodes in the solar corona}

\author[Tomin J. et al.]{
Tomin James,$^{1}$\thanks{E-mail: tomin.james@students.iiserpune.ac.in}
Prasad Subramanian,$^{1,2}$
Eduard P Kontar$^{3}$
\\
$^{1}$Indian Institute of Science Education and Research, Pune, 411008, India\\
$^{2}$Centre for Excellence in Space Sciences, India (http://www.cessi.in)\\
$^{3}$ School of Physics and Astronomy,University of Glasgow,Glasgow, G12 8QQ, UK\\
}

\date{Accepted XXX. Received YYY; in original form ZZZ}

\pubyear{2017}

\begin{document}
\label{firstpage}
\pagerange{\pageref{firstpage}--\pageref{lastpage}}
\maketitle

\begin{abstract}
We study the energetics of nonthermal electrons produced in small acceleration episodes in the solar corona.
We carried out an extensive survey spanning 2004--2015 and shortlisted 6 impulsive electron events detected at 1 AU that were not associated 
with large solar flares(GOES soft x-ray class $>$ C1) or with coronal mass ejections. Each of these events had weak, but detectable hard Xray (HXR) emission near the west
limb, and were associated with interplanetary type III bursts. In some respects, these events seem like weak counterparts of ``cold/tenuous'' flares. The energy carried by the HXR producing 
electron population was $\approx 10^{23}$ -- $10^{25}$ erg, while that in the corresponding population detected at 1 AU was
$\approx 10^{24}$--$10^{25}$ erg. The number of electrons that escape the coronal acceleration site and
reach 1 AU constitute 6 \% to 148 \% of those that precipitate downwards to produce thick target HXR emission.
%These findings are expected to aid our understanding of nanoflare-like events that could contribute towards heating the corona.
 %We report the results of an extensive survey covering a decade carried out
%to find events which has no association with a flare or a CME. We compare the photon spectra of such events 
%observed by RHESSI to the electron spectra observed near earth using ACE/EPAM. We estimate the energy involved in 
%these small electron acceleration episodes in the solar corona which has no association with a flare
%or a CME. For this we rely on the in-situ particle data from ACE/EPAM and the HXR data from RHESSI. Thus
%we carry out a more rigourous analysis of energy released in such small acceleration events compared to relying
%just on radiation data. Further the photon spectral index $\gamma$ and electron spectral index $\delta$ were shown
%to have positive correlation indicating a common physical process behind the production of these
%accelerated electrons.
\end{abstract}


%%%%%%%%%%%%%%%%%%%%%%%%%%%%%%%%%%%%%%%%%%%%%%%%%%

%%%%%%%%%%%%%%%%% BODY OF PAPER %%%%%%%%%%%%%%%%%%

\section{Introduction}

There is ample evidence for acceleration processes in the solar corona that result in
nonthermal particle distributions. These include the radiative signatures and direct particle detections in large
eruptive events that result in flares and coronal mass ejections (CMEs) \citep{Zharkova_2011,Vilmer_2012,Asch2012,Miller_1997,kontar_2011,Holman_2011,Aschwanden_2017}. 
The underlying driver for such particle acceleration events is generally understood to be the excess energy stored in stressed magnetic fields that is released
via the process of magnetic reconnection. While this broad picture has been accepted for a while, it is only recently that observations have started to reveal
some details \citep{Kontar_2017} and simulations have started to
establish the details of particle acceleration in reconnection regions 
(e.g., \citep{Vlahos_2016,Arzner_2004,Dahlin_2015} and references therein).
On the other hand, the possibility of small, ubiquitous reconnection events accelerating electrons and giving rise to the so-called nanoflares
\citep{Parker_1988} has gained considerable momentum as a candidate for
coronal heating (e.g., \citep{Klimchuk_2015,Barnes_2016} and references therein).Recent simulations have demonstrated the spontaneous
development of current sheets with
a high filling factor, even away from magnetic nulls (\citep{Kumar_2015,Kumar_2016}); 
these current sheets can serve as potential sites for small electron acceleration events.
However, since these nanoflares are very small, its very difficult to observe them directly 
(e.g. \citep{Testa_2013,Joulin_2016}), and one can only make indirect 
inferences about them. There are only a few claims in the literature regarding detection of nanoflare (or even smaller) energy releases 
at radio wavelengths \citep{Mercier_1997,Ramesh_2012}. In most instances,
the observation and interpretations are concerned only with the radiative signatures arising from the electrons which are accelerated by the 
reconnection episodes. 
%In the causative chain that starts with reconnection, leading to electron acceleration and culminating in the observed radiation,
%they focus only on the last stage.

We are concerned with the second stage of the chain that starts
with reconnection, leading to electron acceleration and culminating in the observed radiation. In this paper, we study the energy budgets and other characteristics of accelerated electrons.   
The electrons responsible for these events are typically accelerated in the corona; some of them precipitate downwards into denser layers of the solar atmosphere to 
produce Hard X-ray (HXR) emission, while some find access to open field lines and travel outwards, often being detected as electron spikes at
1 AU by spacecraft such as WIND, ACE and STEREO \citep{Klassen_2012}. We only study impulsive electron events detected at 1 AU that are unaccompanied by soft 
X-ray flares and coronal mass ejections, e.g. \citep{simnett2005}, so that we can be reasonably sure that the energy releases involved are indeed small. 
Electron beams travelling outward through the corona have well established radio signatures, called type III bursts in the corona 
e.g., \citep{Hilaire_2012}, and IP type III bursts in the interplanetary medium 
\citep{Krupar_2014,Krupar_2015}. There is recent evidence for very weak type III bursts 
\citep{Tun_2015} that could provide interesting information regarding the relatively weak events 
that generate the electron beams responsible for this emission.

One of the most interesting questions that can be answered by our study concerns the fraction of the electrons accelerated in the
coronal site that escape and reach the Earth - as compared to the ones that travel downwards to the chromosphere and produce HXR emission. Other associated
quantities include the power and energy carried by the escaping electrons, and a comparison of the corresponding quantities for the HXR emitting electrons.

%\section{Observations}
%Parker(1988) postulated the possibility of tiny magnetic reconnection events
%happening all across the solar corona as one of the sources of coronal heating. They are called nanoflares
%as the energy involved ($\sim 10^{24}$ ergs) is nine orders of magnitude smaller than the biggest of flares.
%The free energy released goes into heating thermal and non-thermal particles. Studies have shown that most  
%fraction of free energy goes into non-thermal paritcles (add citations). Part of the energy that goes into 
%non-thermal particles are carried away by escaping electrons from the accelration site which then travel through
%open-magnetic field lines are observed near earth. We combine this in-situ particle data observed with EPAM instrument
%on board ACE spacecraft and HXR data from RHESSI spacecraft to better understand the fraction of energy input into
%non-thermal particles.

%{\color{blue} TOMIN - START HERE}

\section{Event Shortlisting}
\label{sec:PES} % used for referring to this section from elsewhere
Our primary focus is on small impulsive electron events observed in-situ at 1 AU by the EPAM detector aboard the ACE spacecraft. 
We carried out an extensive survey using ACE/EPAM data spanning the years 2004--2015, which covered the maxima of solar cycles 23 and 24 and 
the intervening minimum.
ACE/EPAM is a suite of five instruments, three of which respond to electrons. The magnetically deflected electron detector, 
LEMS30 has four energy channels
from 38-315 keV. These channels are named DE1(38-54 keV), DE2(53-103), DE3(103-173 keV) and DE4(175-315 keV). We only include events that
were clearly detected 
in at least the first three channels (DE1 - DE3).
We furthermore require that these impulsive electron events are 
\begin{itemize}
\item
not associated with large GOES soft X-ray flares or CMEs, 
\item
associated with interplanetary type III bursts 
\item
associated with reliable west limb signatures in RHESSI hard X-ray data. 
\end{itemize} 
We also searched the Solar Geophysical Data database (for events prior to 2009), USAF-RSTN and e-CALLISTO data for possible association with microwave bursts,
which would indicate chromospheric activity.
None of the events we shortlisted were associated with microwave bursts. The lack of chromospheric acitivity suggests a coronal origin for these events,
as does the steepness of the energy spectra e.g., \citep{potter1980}.  
%We next elaborate on the shortlisting procedures.

\subsection{Onset times at 1 AU}
Before we elaborate on the shortlisting procedures, we mention the manner in which we calculate the event onset times at 1 AU. The impulsive electron events are detected at 1 AU in 4 energy channels
by ACE/EPAM. Each of these energy channels can be associated with an average electron speed $v(E)$. 
In order to accurately measure the energy spectrum of the in-situ electrons, we have excluded events which were not clearly resolved in the first three
energy channels of ACE/EPAM. The event onset at the Sun ($t_{sun}$) is taken to correspond to the peak of the HXR flux in the RHESSI 12-25 keV channel. 
%This particular channel was chosen to have a consistent
%measure for all the events, as the counts were weak for most events.
Following \citet{krucker1999}, we calculate the event onset time at 1 AU via
\begin{equation}
t_{1AU} = t_{sun} + \frac{L}{v(E)} \, ,
\label{eq1}
\end{equation} 
where $L$ is the path length (taken to be equal to the average Parker spiral length of 1.2 AU) and $v(E)$ is the particle speed
corresponding to the average energy of the appropriate ACE/EPAM channel \citep{krucker1999,krucker2007,potter1980}. 
The expected arrival times for each energy channel are plotted as colour coded dashed lines in the ACE/EPAM panel of Figure~\ref{fig:stacked}.

\subsection{Shortlist 1: Non-association with CMEs and GOES flares}
Near-relativistic electrons associated with CMEs are typically released around 20 min after the launch of a typical CME, by which time 
the CME front is around 1.5 - 3 $R_{\odot}$ from the Sun \citep{sim2002}. Using CME onset times from the LASCO CME catalog (http://cdaw.gsfc.nasa.gov/CME list/), 
we ensure that the impulsive electron 
events we shortlist are not associated with CMEs observed between $t_{sun}$ and the event onset at ACE $t_{1AU}$. 
%Using the electron injection time determined
%previously and CME height-time plot, the relative
%position of CME at the time of electron injection can be reliably find out.
We further note that CME-associated shock driven electron events typically have 
rise times ranging from tens of hours at ACE/EPAM, whereas the events we study have rise and decay times of a few tens of minutes, with an average 
duration of 20 minutes in the DE2 channel.

We also make sure that there are no GOES soft X-ray flares larger than the C1 class within 5 minutes of the event onset time $t_{sun}$. 
%We require that none of our flares has any association with X-ray brightnenings
%greater than C1 class or $10^{-06} Watts/m^{2}$ in absolute magnitude.
%Any X-ray activity below GOES C1 class
%were omitted as such low SXR flares are not capable of accelerating
%significant amount of electrons.
This shortlisting procedure yields a database of 18 events
from 2004 to 2015, the details of which are summarized in Table \ref{tab:eventlist}.  This includes seven out of the nine events reported by \citep{simnett2005}.
%The onset time at the Sun is the quantity $t_{sun}$ (Eq~\ref{eq1}), which gives the time when the impulsive electrons were produced at the Sun. 
The column 
titled ``Onset at GOES'' gives the time when one can discern a rise in the GOES soft Xray flux. As the entries under the column titled 
``GOES flux level'' indicate,
none of the GOES soft Xray enhancements are above the C1.1 level. The column ``Onset at ACE-DE4'' indicates the time of the peak value at the 
DE4 energy channel(mean energy 275 keV) in ACE/EPAM. 
%All our events have a sharp rise and decay times, with the average duration of an event to be $\sim$ 35 minutes.
%The time of flight estimated from the onset times at Sun and the spacecraft is $\sim$ 15 minutes, suggesting a scatter free propagation through interplanetary
%space
\begin{figure*}
 
\centering
\begin{tabular}{cc}
 \hspace*{-1.2cm} 
  \subfloat[]{\includegraphics[scale=0.63]{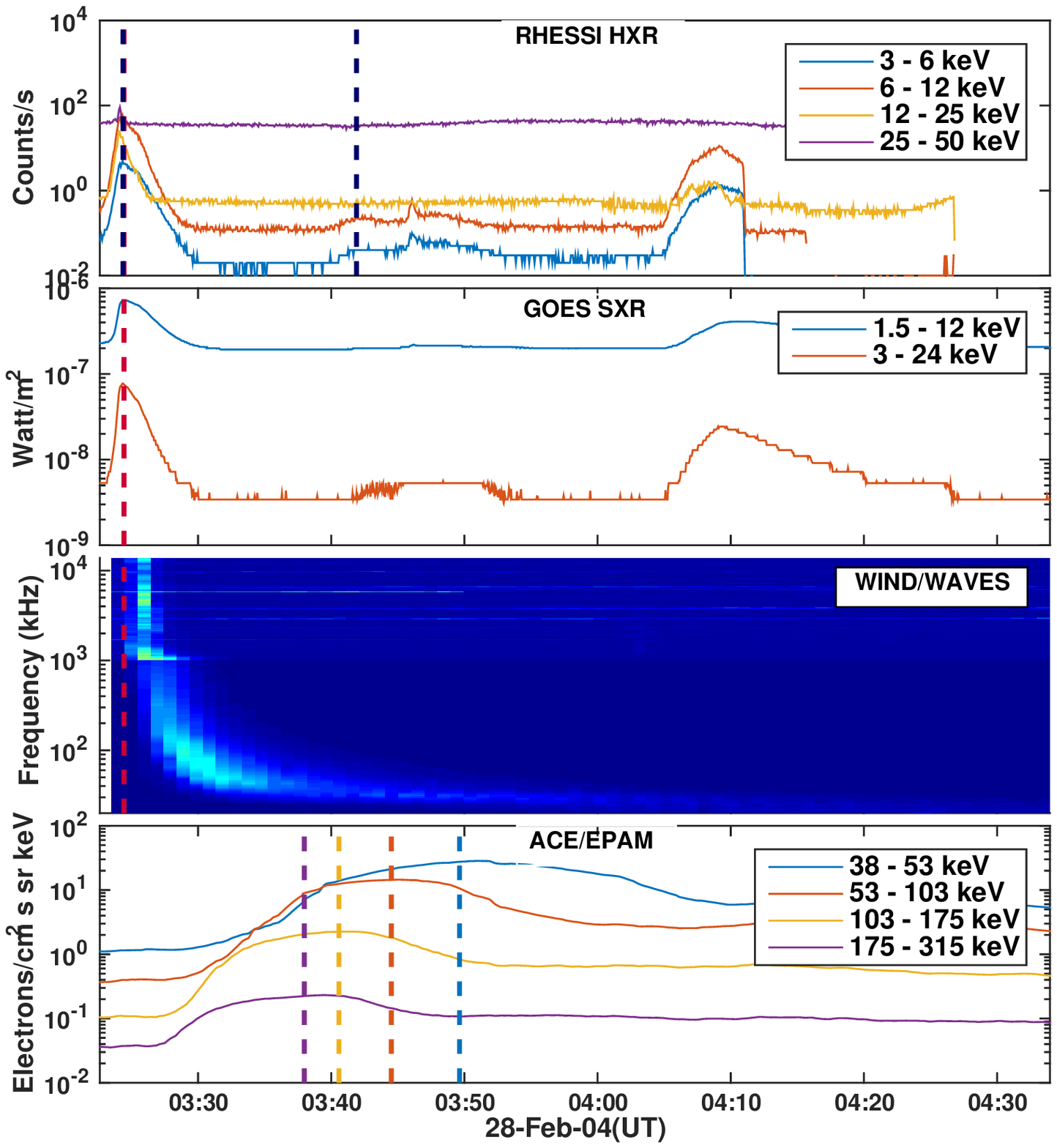}} & 
  \subfloat[]{\includegraphics[scale=0.63]{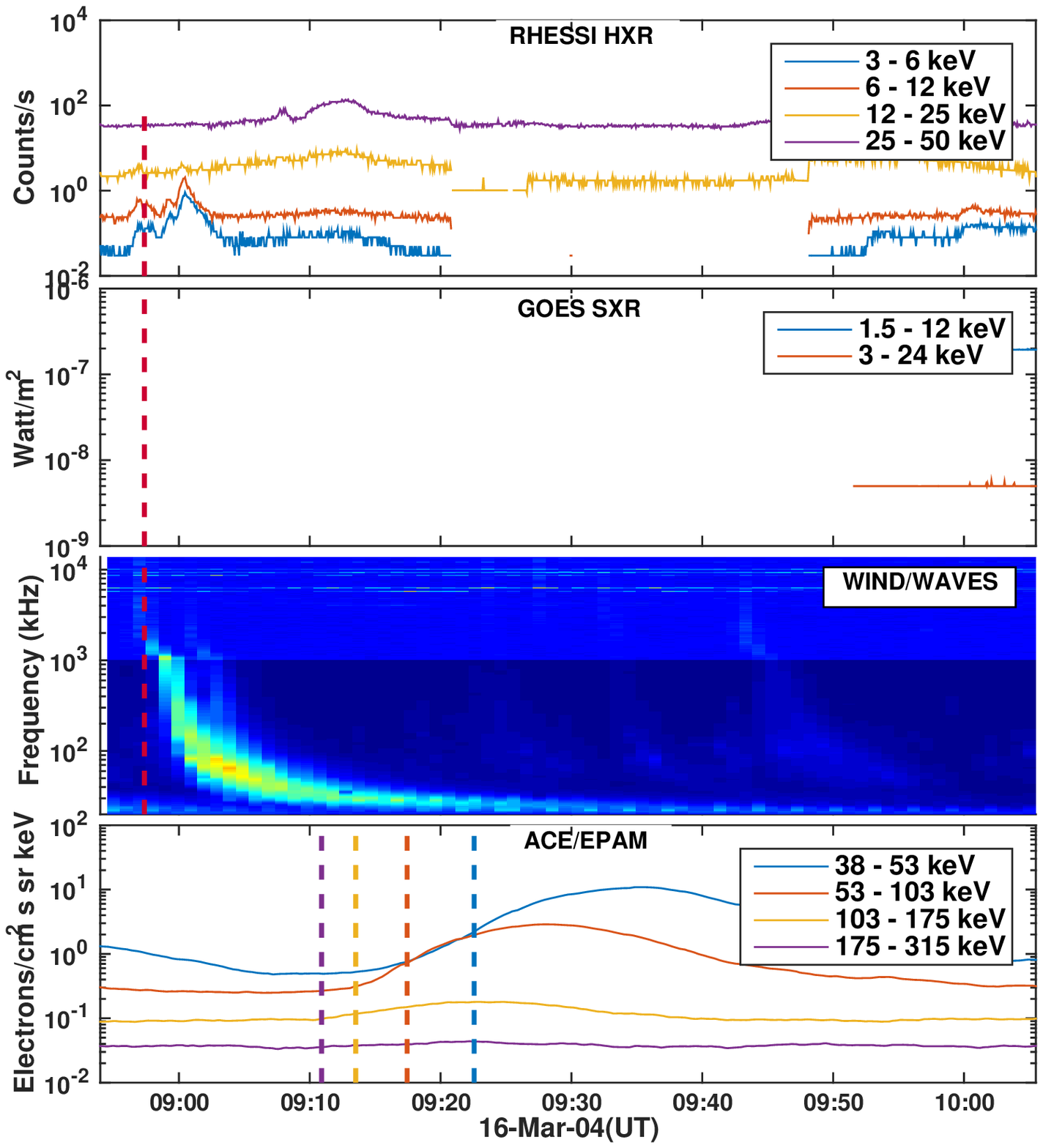}} \\ 
  \hspace*{-1.2cm} 
  \subfloat[]{\includegraphics[scale=0.63]{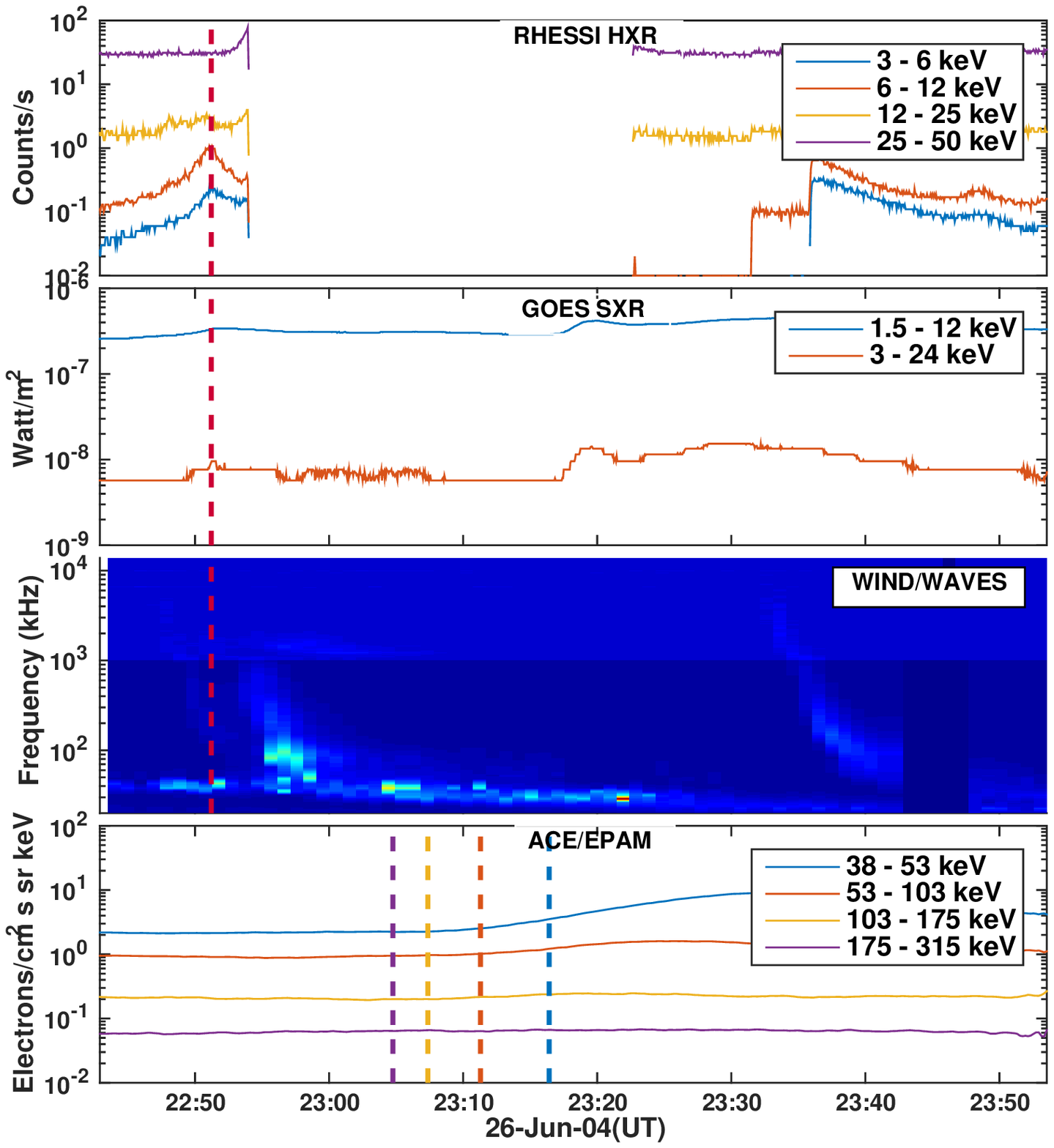}} &
   \subfloat[]{\includegraphics[scale=0.63]{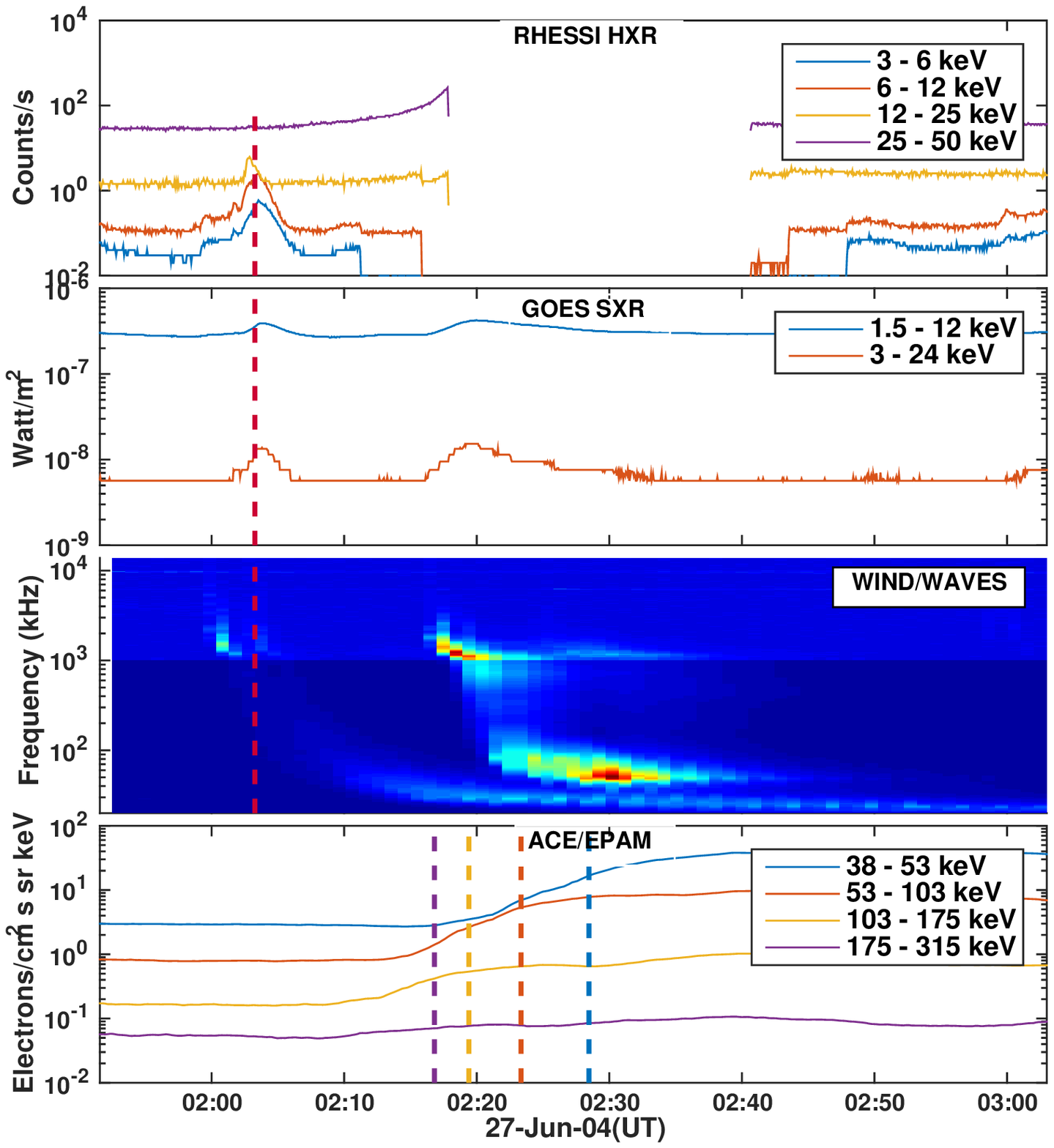}} \\
   
  \end{tabular} 
  \phantomcaption
  \label{fig:stacked}
  \end{figure*}
  
  \begin{figure*}
  \centering
  \begin{tabular}{@{}cc@{}}
   \ContinuedFloat
    \phantomcaption
   \hspace*{-1cm} 
  \subfloat[]{\includegraphics[scale=0.63]{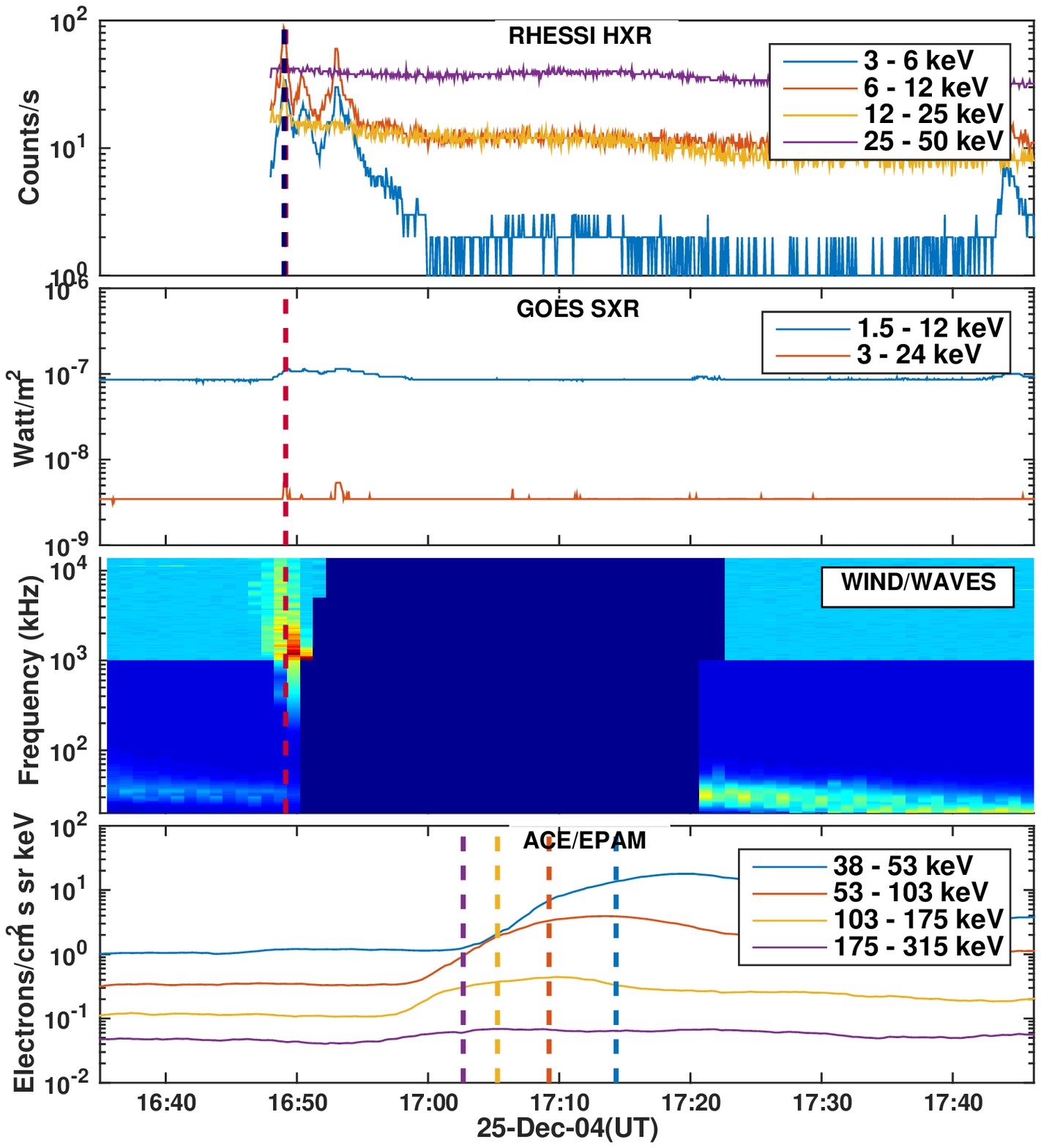}} &
  \subfloat[]{\includegraphics[scale=0.63]{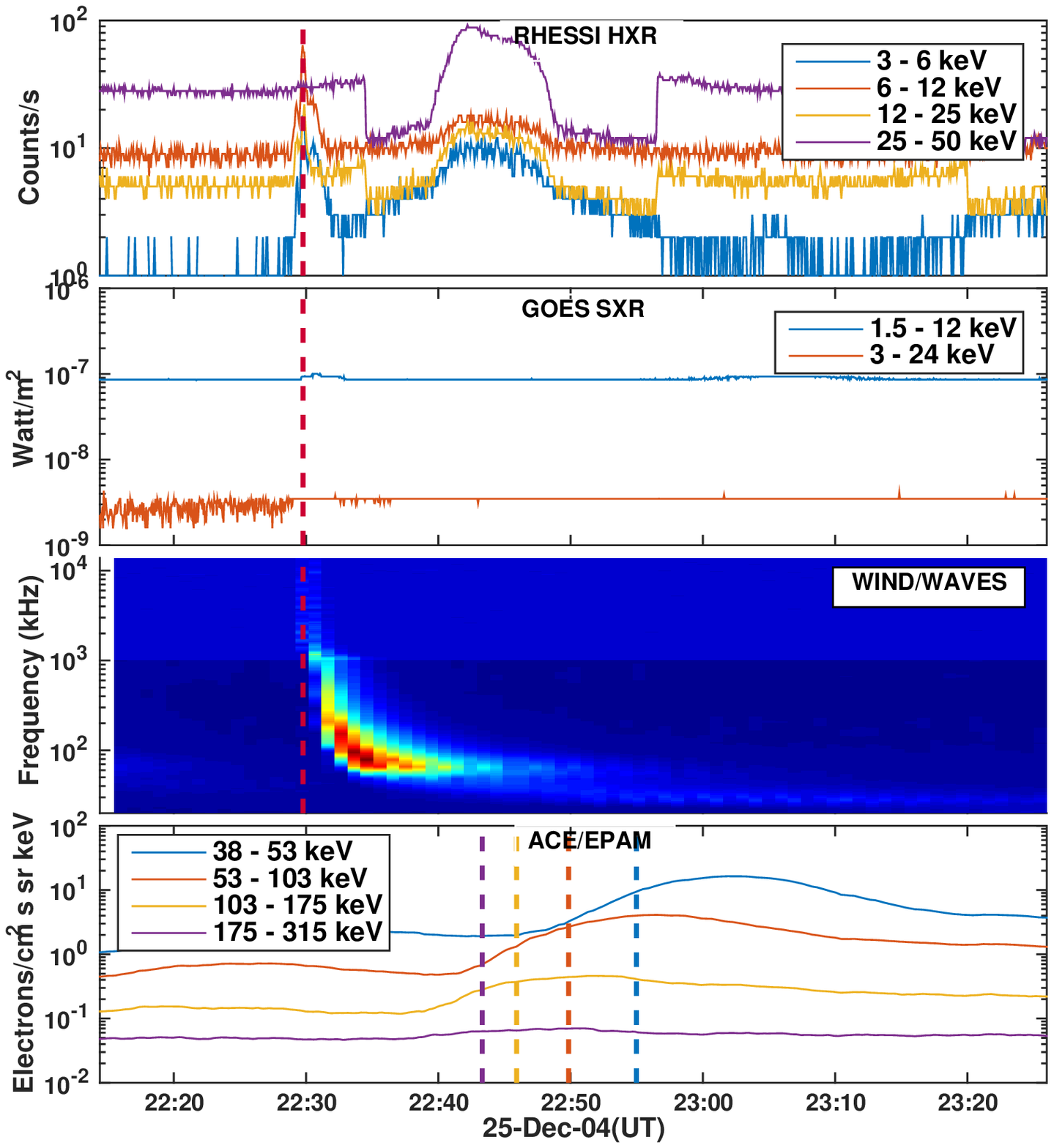}} \\

  \end{tabular}
  \caption{Stacked plots for each of the six shortlisted events. The top panel shows the corrected RHESSI HXR photon counts. The dark red
   dashed line depicts the quantity $t_{sun}$, which corresponds to the peak of the HXR emission in the 12-25 keV channel. The dark blue dashed line depicts the timing of the HXR flare from the RHESSI flare list, where available.
   Since our shortlisted events are very weak, 
   only two events are found in the RHESSI flare list. The second panel shows the GOES SXR flux. 
   %The dark red line indicates
   %the onset time of the HXR flare at the sun. 
   The third panel is the
   WIND/WAVES dynamic spectrum, showing the interplanetary type III burst associated with each event.
   The bottom panel shows the time evolution of the electron flux in the four ACE/EPAM energy channels. The thick dashed lines correspond to the expected arrival times of the 
   electrons in the different energy channels, assuming that they are released at $t_{sun}$ and travel a distance of 1 AU.}
   
\end{figure*}

\begin{table*}
\caption{Impulsive electron events detected in-situ at 1 AU: first shortlist }
\label{Q-sun events}
\begin{threeparttable}
\begin{tabular}{lccccc}     % define the column alignment
                           % l: left, c: center, r: right
\hline                   % horizontal line
 Event Date        & Onset at           & Onset at        &GOES flux      &     Type III onset at             &SEP onset                           \\
                   & Sun ($t_{sun}$)           & GOES       &level          &     WIND/WAVES      & at ACE-DE4                        \\
                   & (UT)                   &(UT)           &                 &(UT)                   &(UT)                  \\
\hline
Feb 28,2004\tnote{*}         & 03.24              & 03.24            &B6.6           &     03.25                  &03.36\\ %windisthere
Mar 16,2004\tnote{*}         & 08.56              & no-data            &C0.1           &     08.57                   &09.12                           \\
June 26,2004                 & 20.50              & nil             &B2.1           &     20.51                  &21.13  \\
June 26,2004\tnote{*}        & 22.50              & 22.50           &B2.2           &     22.51                   &23.02            \\
June 27,2004\tnote{*}        & 02.03              & 02.02           &B1.1           &     02.03                   &02.17 \\
June 27,2004                 & 04.59              & 05.01           &B1.1           &     05.01                   &05.19\\
June 27,2004                 & 13.02              & 13.01           &C0.0           &     13.02                  &13.21\\
June 27,2004                 & 14.59              & nil             &B1.1           &     15.01                   &15.26\\
June 27,2004                 & 17.43              &  nil            &B1.1           &     17.43                   &18.10                       \\
Dec  25,2004\tnote{*}        & 16.49              &  16.50          &B0.0           &     16.49                 &17.03                       \\
Dec  25,2004\tnote{*}        & 22.29              & 22.29           &B0.1           &     22.29                  &22.43                            \\
Mar 16,2005                  & 23.03              & 23.02           &B1.3           &     23.02                   &23.19 \\
Jan 13, 2007                 & 15.10              &15.11            &B2.2           &     15.12                   &15.21 \\
Feb 18, 2010                 & 18.58              &18.56            &A9.1           &     18.59                   &19.14\\
Mar 28, 2014	             &20.57		&20.58		&C0.0		&20.58				&21.18\\%onlyfrom40keV
Jan 20, 2015	             &09.48		&09.49		&B9.1		&09.50				&10.05\\%nowind
May 14,2015	             &05.08		&05.10		&C0.6		&05.10				&05.23\\%nowind
May 14,2015	             &07.28		&07.29		&C1.0		&07.29				&07.40\\ %nowind

\hline
\end{tabular}
 \begin{tablenotes}
  \item[*] Events in the final shortlist.
  \end{tablenotes}
  \end{threeparttable}
\label{tab:eventlist}
\end{table*}

\subsection{Shortlist 2: Association with IP type III bursts}
We would expect the escaping electrons to excite interplanetary type III bursts.
%Since we presume that the impulsive electron events detected at the Earth originate in the solar corona and travel through the heliosphere, 
%it is natural to look for associations with interplanetary type III bursts. Electron beams passing through solar corona excite radiation at local plasma
%frequency causing type III burst. These bursts act as an indicator for the passage
%of accelerated electrons. 
We use data from the WAVES RAD1 (20 - 1040 kHz) and RAD2 (1.075 - 13.825 MHz) detectors aboard the WIND spacecraft to search for
interplanetary type III bursts associated with the impulsive electron events detected by ACE/EPAM. We find that all the events have an associated
IP type III burst with onset times within a minute of the onset at the Sun $t_{sun}$. This indicates that electron injection into the interplanetary
space happened during the flaring process and any delay at ACE/EPAM is purely due to propagation effects. This is unlike delayed SEP events, in which
electron injection happens tens of minutes after the flaring process.%We include only events that have IP type III counterparts. 
The column titled ``Onset at WAVES'' in Table \ref{tab:eventlist}  indicates the time of onset of the IP type III burst in the WIND/WAVES RAD2 receiver.
%It may be noted that the type III onset is typically within a couple of minutes of $t_{sun}$ (Eq~\ref{eq1}), indicating near instantaneous injection
%of particles into interplanetary space following the acceleration episode. %The results are tabulated in Table 1.
%The events were also crosschecked with Solar Geophysical Data 
%for any bursts or microwave activity. Solar Geophysical Data reports stopped publishing in 2009. Hence for event search post-2009 we used USAF-RSTN
%data.
%This results in 18 events that 
%are shown in Table 1. 

   % \begin{figure}    %%%%%%%%%%%%%%%%%% FIGURE 2 
  % \centerline{\includegraphics[width=\columnwidth,height=7cm]{onset_time.jpeg}
%              }
 %             \caption{The electron onset time as observed from 1 AU. Time axis intercept gives the onset time at the sun. The red line is the
            %  linear fit. The slope of the graph gives the path length in AU.
%                      }
 %  \label{onset_time}
 %  \end{figure}

%\begin{equation}
  %  x=\frac{-b\pm\sqrt{b^2-4ac}}{2a}.
	%\label{eq:quadratic}
%\end{equation}

%Refer back to them as e.g. equation~(\ref{eq:quadratic}).
\subsection{Shortlist 3: Association with HXR emission}
We further require that the events we analyze have enough HXR photon counts for carrying out spectroscopic and imaging anaylsis using RHESSI data. We require that the the RHESSI signatures are located near the west limb, which would enable better magnetic connectivity to the Earth for escaping electrons (Figure \ref{fig:pixon} 
and Table \ref{tab:spectrum}). This defines our 
third and final shortlist, comprising 6 events, which are listed in Table \ref{tab:spectrum}. We note that all but one of the events (Feb 28 2004) have very few photon counts above 35 keV. Since most of the events are very weak compared to standard RHESSI flares only two of them have been included in the RHESSI flare list
(${\rm http://hesperia.gsfc.nasa.gov/hessidata/dbase/hessi_{-}flare_{-}list.txt}$).

 \section{Data Analysis}
 
 The general picture we have in mind is as follows: electrons are accelerated in the corona during episodes of magnetic reconnection.
Some of these accelerated electrons propagate downwards towards the Sun and produce HXR emission via the thick-target bremsstrahlung process.
Some find access to open magnetic field lines, manifesting as impulsive electron events detected in-situ by ACE/EPAM and producing interplanetary type 
III bursts on the way.

The stacked plots shown in figure~\ref{fig:stacked} depict the observational signatures of this chain of events.  The red dashed line denotes the event onset time at the Sun ($t_{sun}$).  The topmost panel depicts the HXR photon counts measured by the RHESSI detectors.
The second panel shows the GOES soft Xray flux. There is a small enhancement of the GOES soft Xray flux for each event, although the level is $<$ C1. The low frequency radio dynamic spectrum recorded by WIND/WAVES is depicted in the third panel; it shows interplanetary type III bursts due to the electrons escaping from the coronal acceleration site. The bottom panel shows the electron flux measured in-situ by ACE/EPAM. The dashed lines show
expected arrival times for the different energy channels computed according to Eq~\ref{eq1}. 
%{\color{red} RIGHT, TOMIN?}

%Figures are referred to as e.g. Fig.~\ref{fig:example_figure}, and tables as
%e.g. Table~\ref{tab:example_table}.

\begin{figure*}
\begin{tabular}{ccc}
  \hspace*{-1.6cm} 
  \subfloat[]{\includegraphics[scale=0.55]{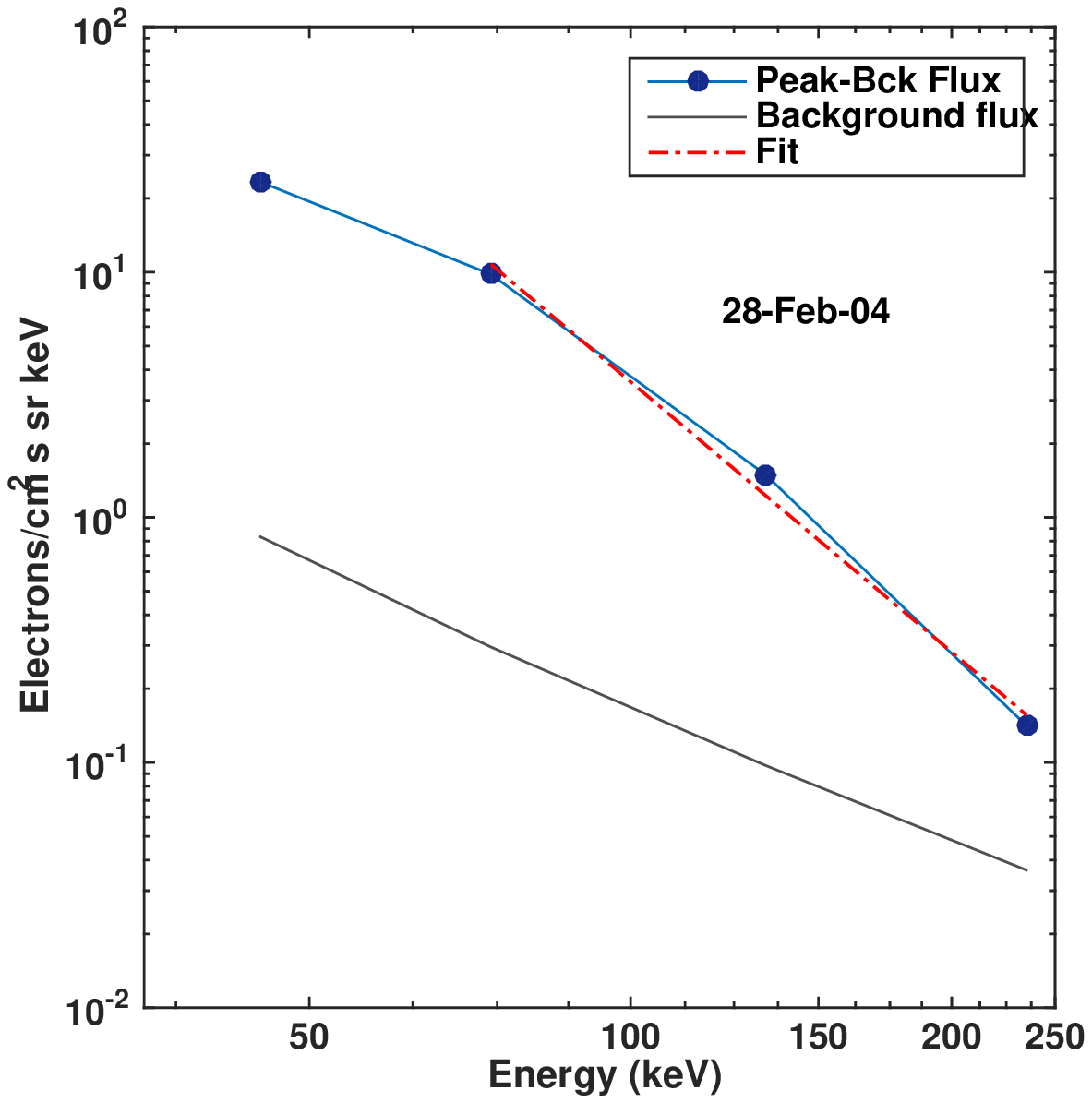}} &
  \hspace*{-1cm}
  \subfloat[]{\includegraphics[scale=0.55]{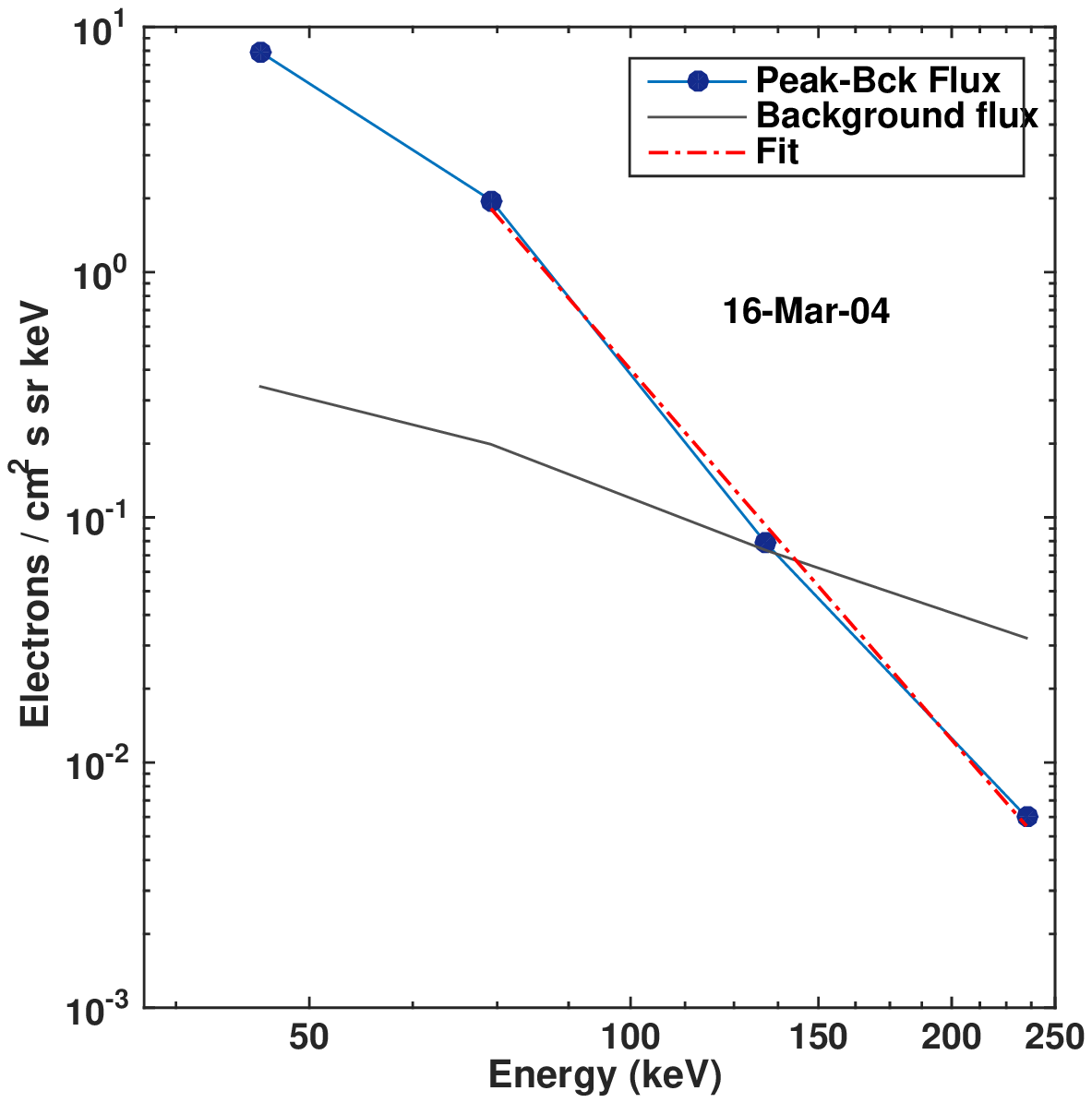}} &
  \hspace*{-0.9cm} 
  \subfloat[]{\includegraphics[scale=0.55]{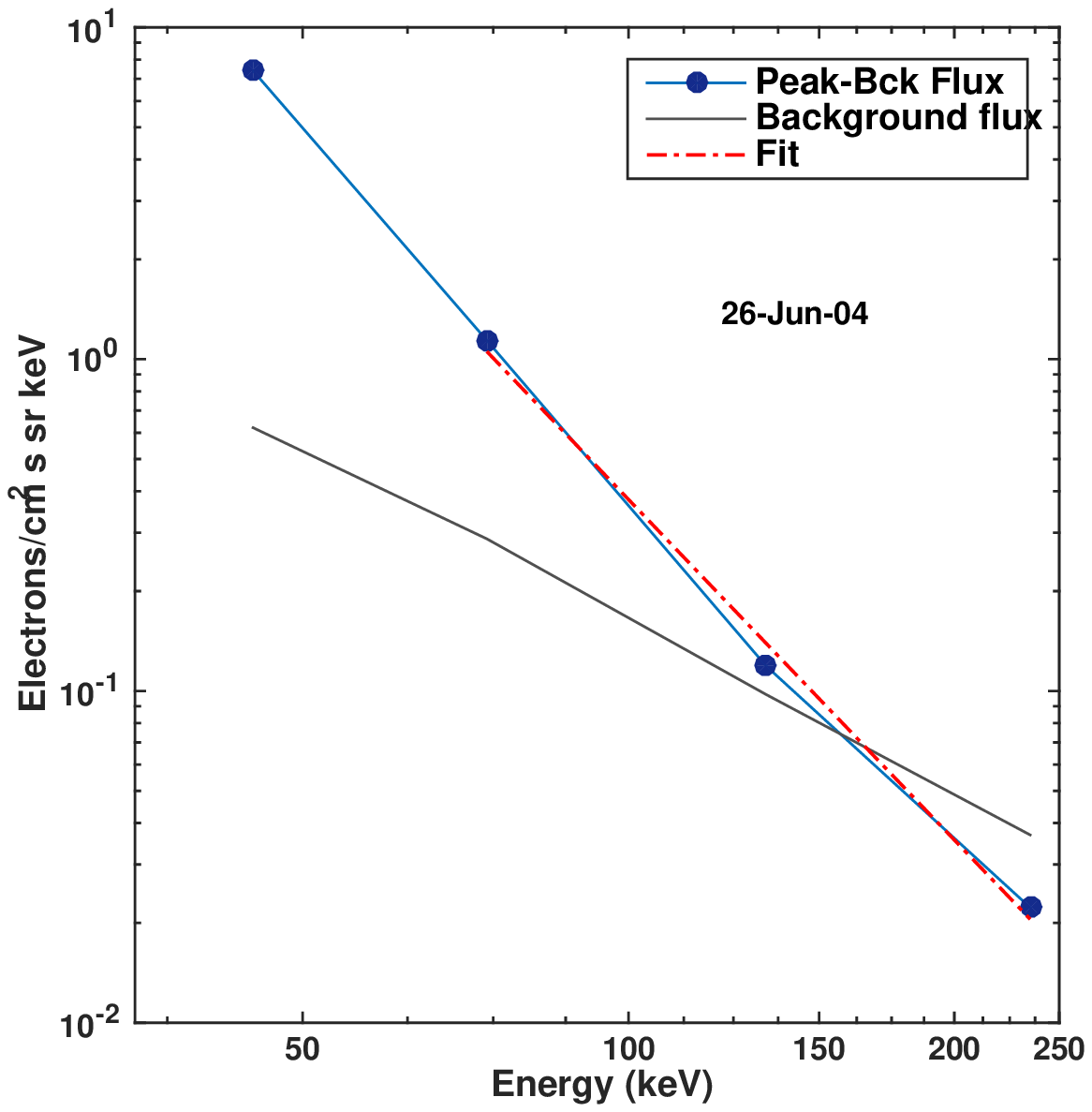}} \\
  
   \hspace*{-1.6cm}
  \subfloat[]{\includegraphics[scale=0.55]{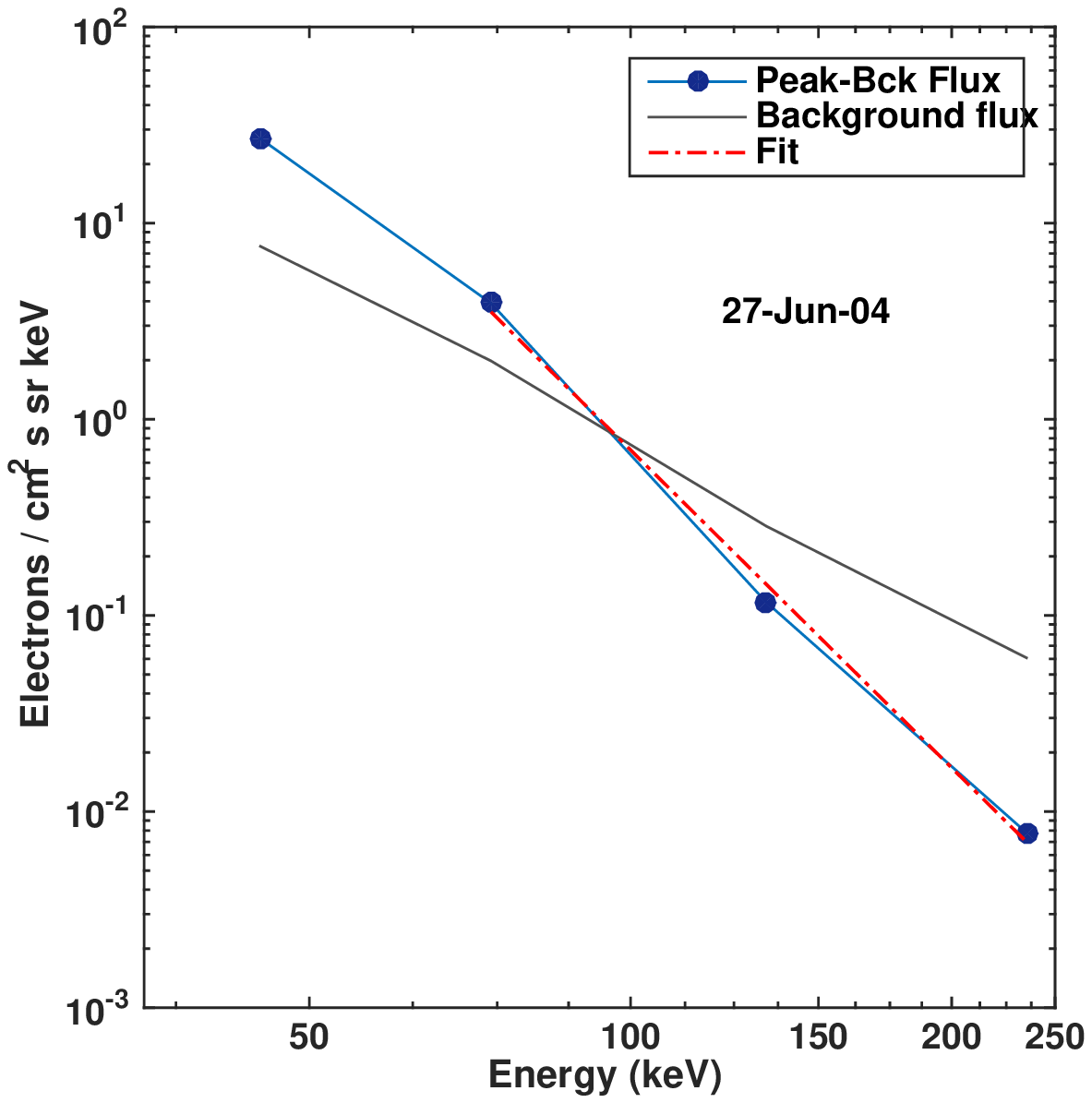}} &  
  \hspace*{-1cm} 
  \subfloat[]{\includegraphics[scale=0.55]{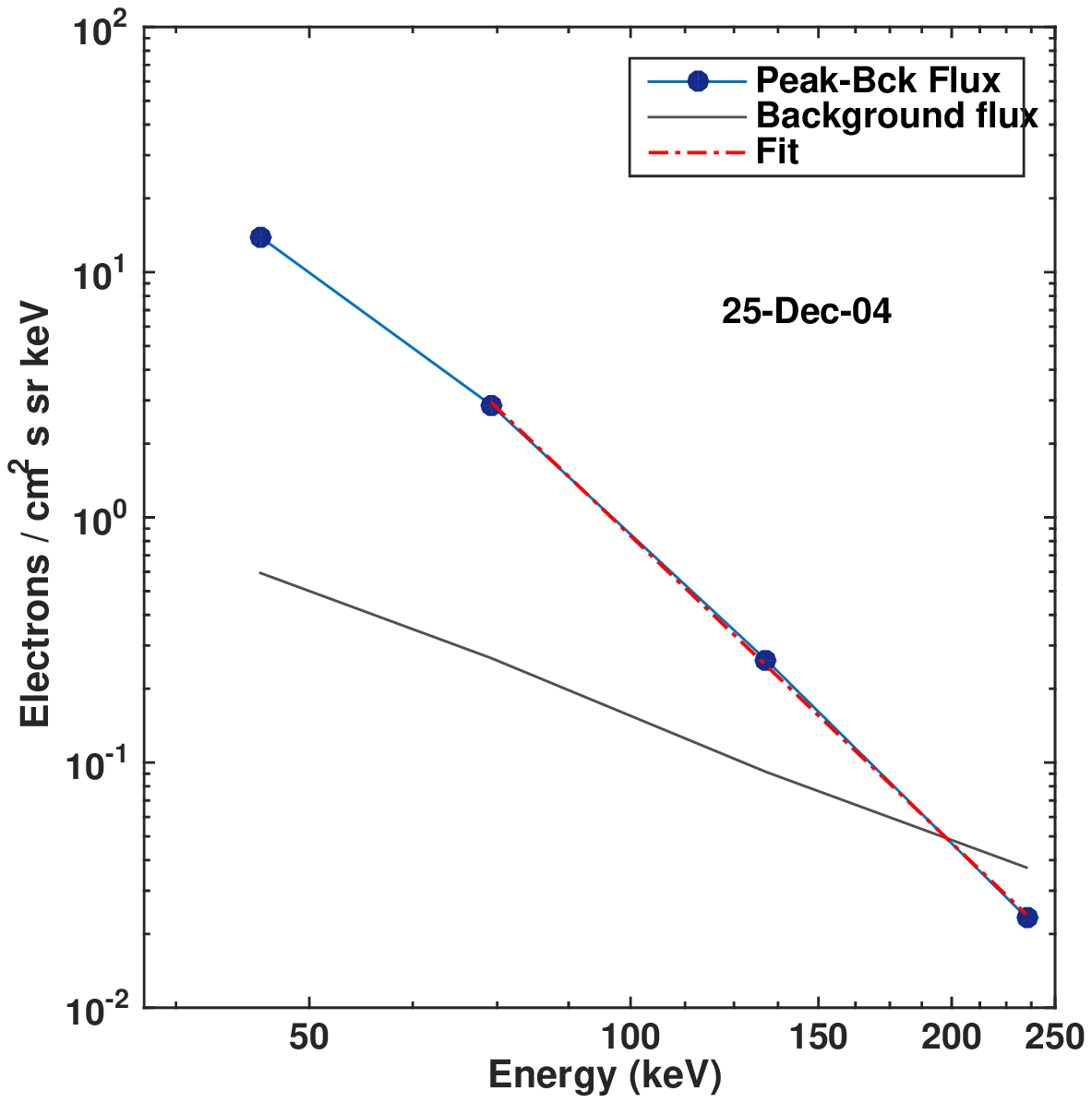}} &
 \hspace*{-0.9cm} 
  \subfloat[]{\includegraphics[scale=0.55]{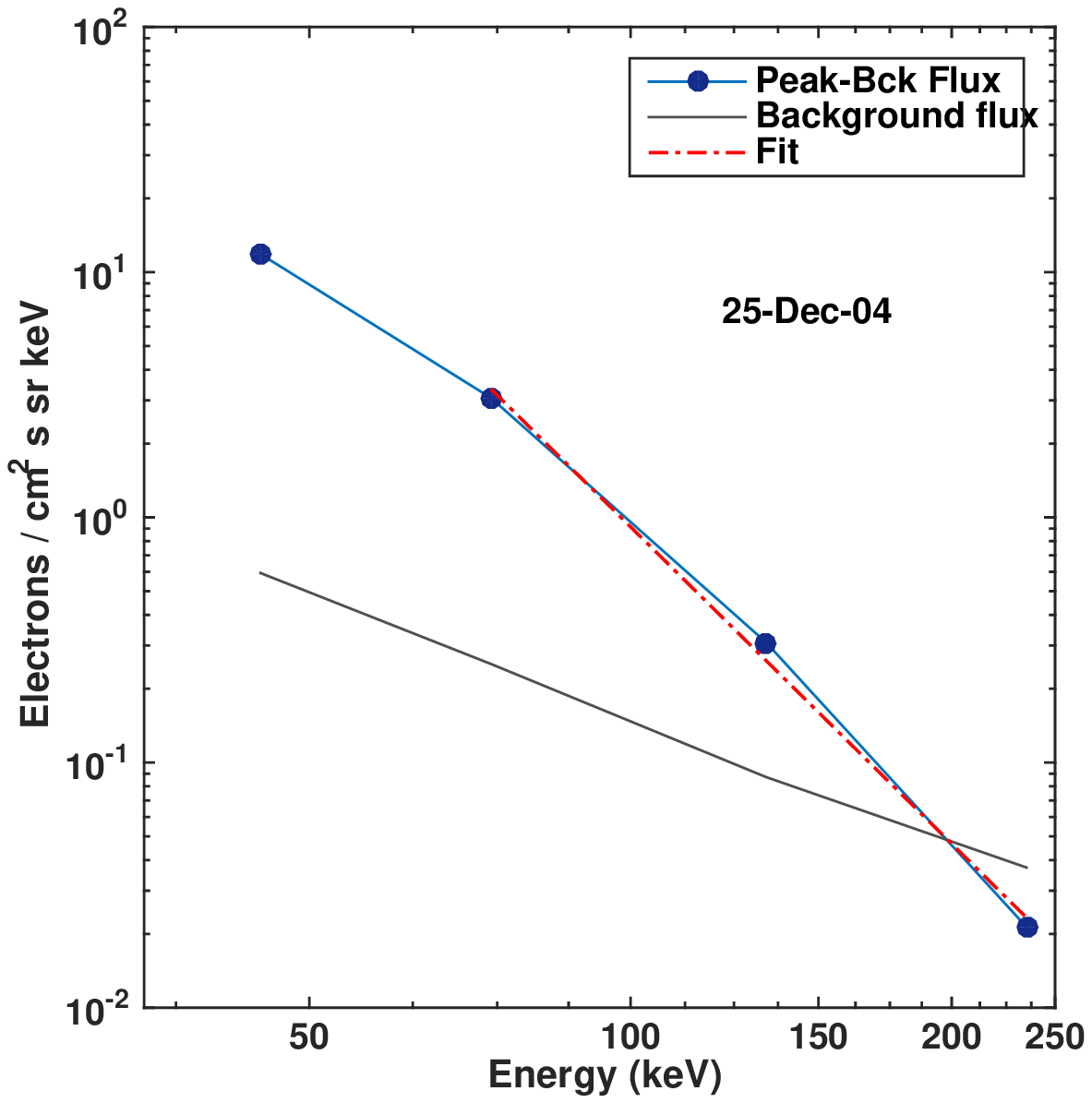}} \\
  \end{tabular}
  \caption{ Fits to the ACE/EPAM electron data for each of the six shortlisted events. The background spectrum was subtracted to the from the spectrum at the event peak.
             We observe a break in the spectrum for all the events at 74 keV. We fit a power law
              to the spectrum above the break. This is depicted by the thick blue line in the graphs. Results are tabulated in Table \ref{tab:spectrum}.}
   \label{fig:delta}
\end{figure*}
 
%In this section we describe how we estimate the energy contained in the electrons detected at 1 AU as well as the HXR producing
%electrons.
\begin{figure*}

\begin{tabular}{ccc}
\hspace*{-1.9cm}
  \subfloat{\includegraphics[scale=0.5]{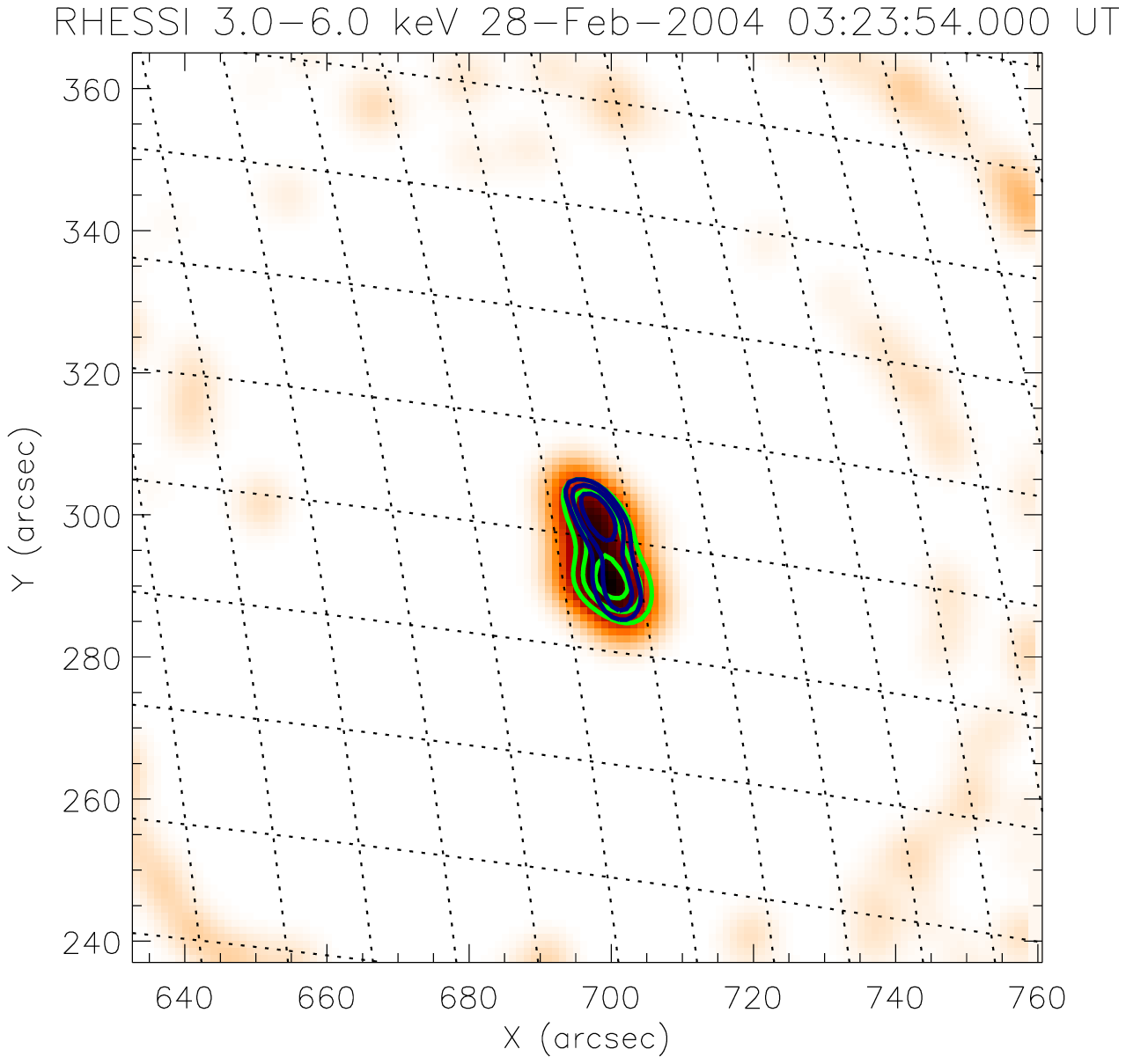}} &
  \hspace*{-1cm}
  \subfloat{\includegraphics[scale=0.5]{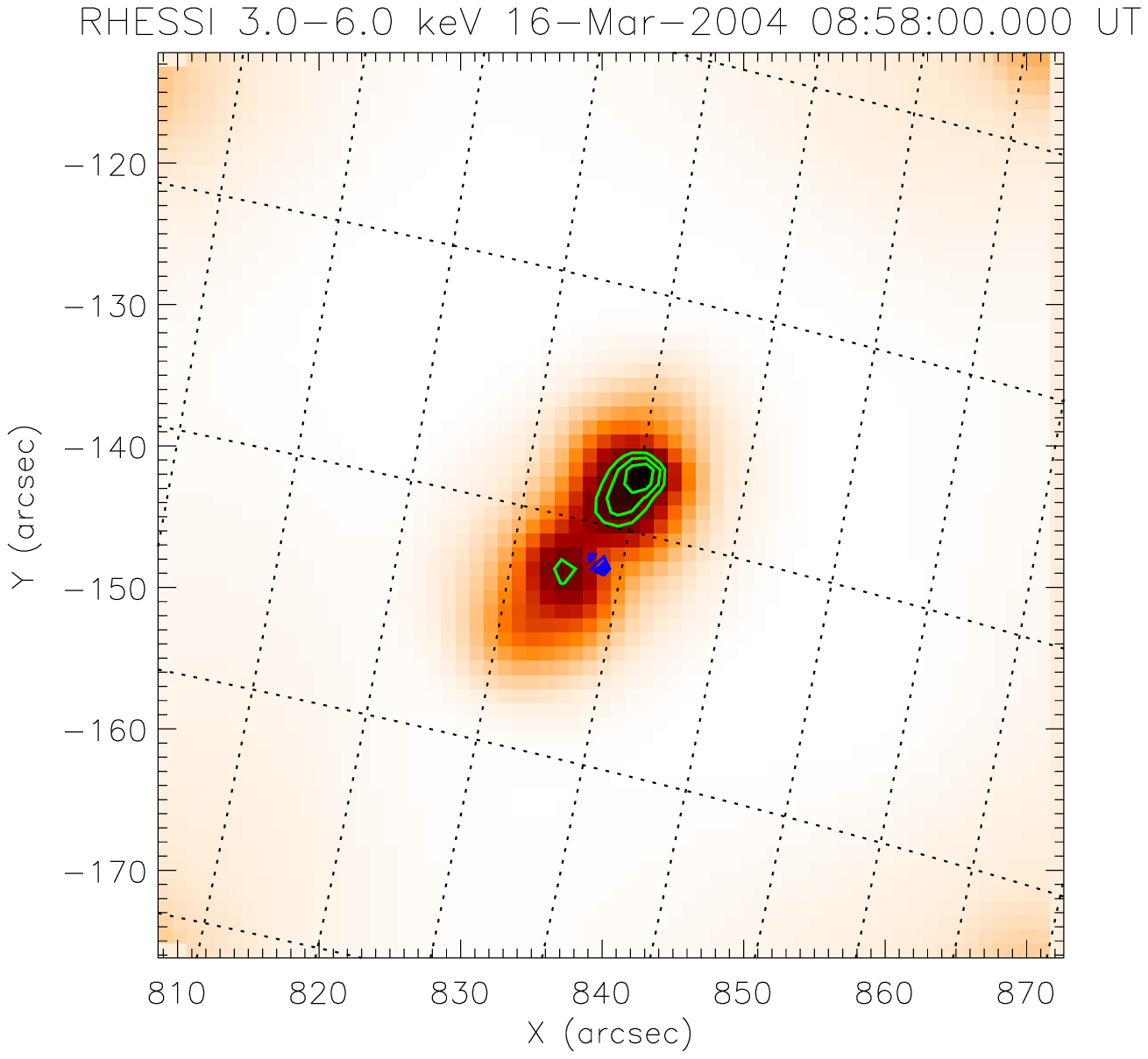}} &
  \hspace*{-1.4cm}
  \subfloat{\includegraphics[scale=0.5]{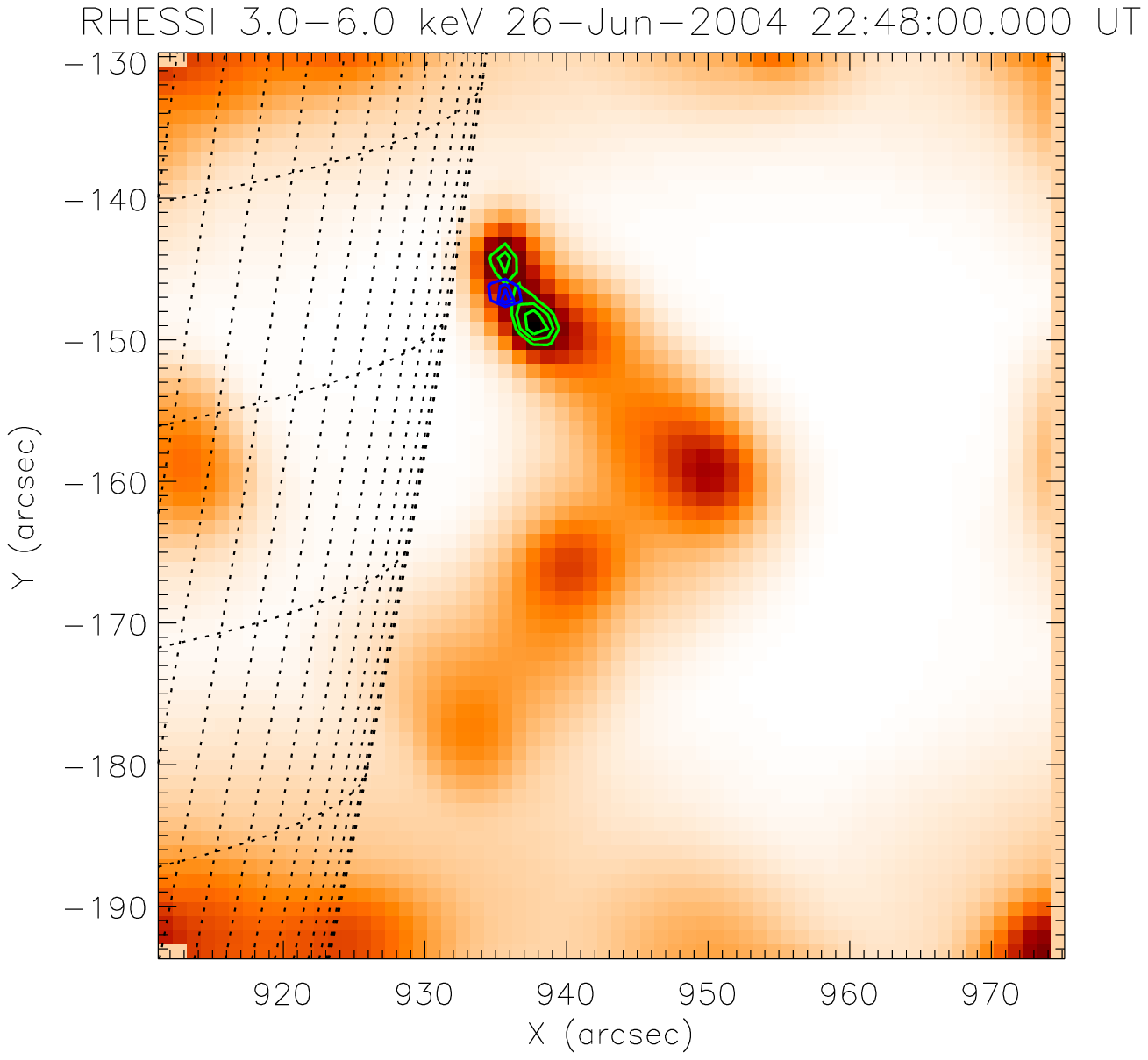}} \\
  \hspace*{-1.9cm}
  \subfloat[]{\includegraphics[scale=0.5]{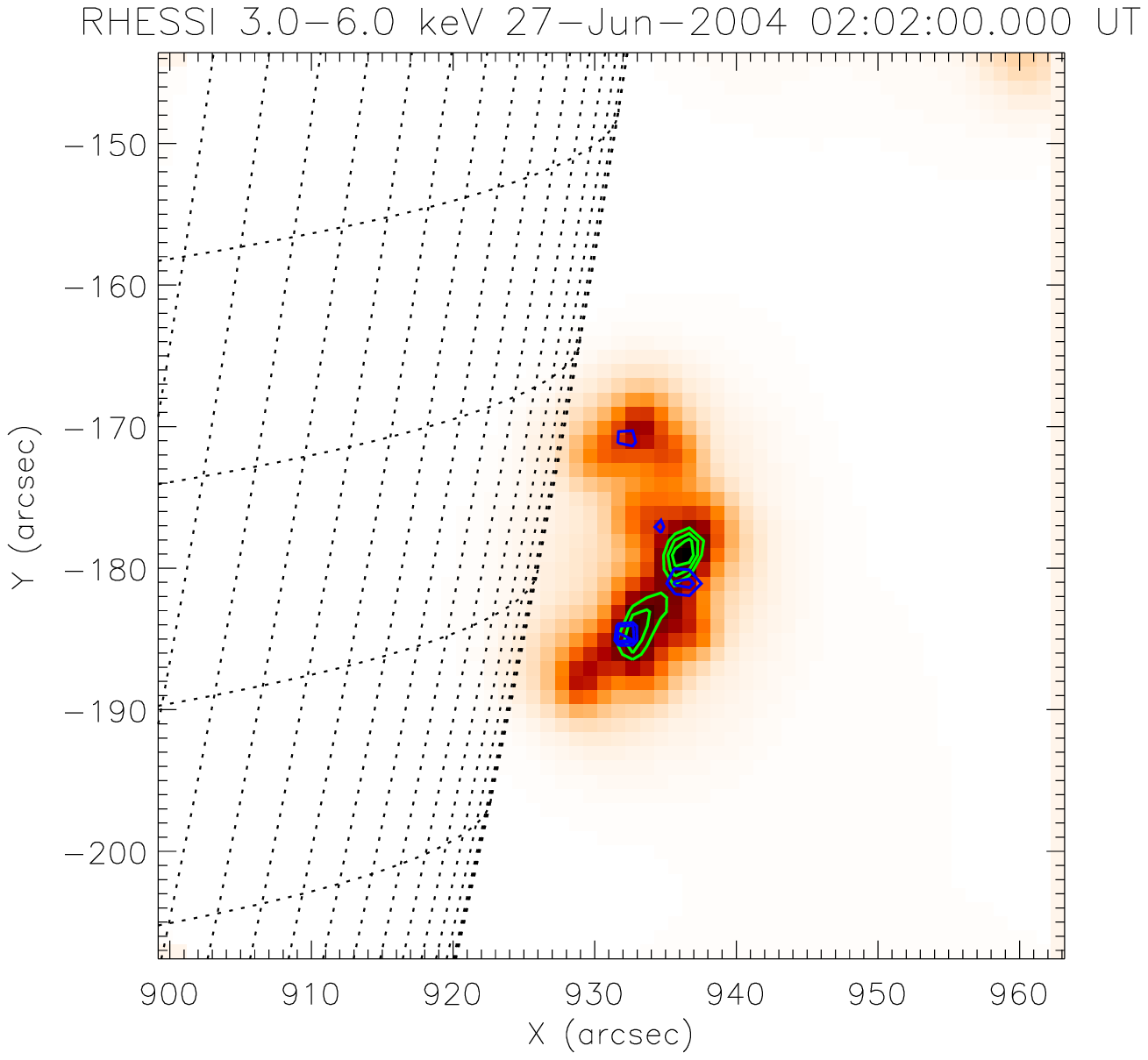}} & 
   \hspace*{-1cm}
  \subfloat[]{\includegraphics[scale=0.5]{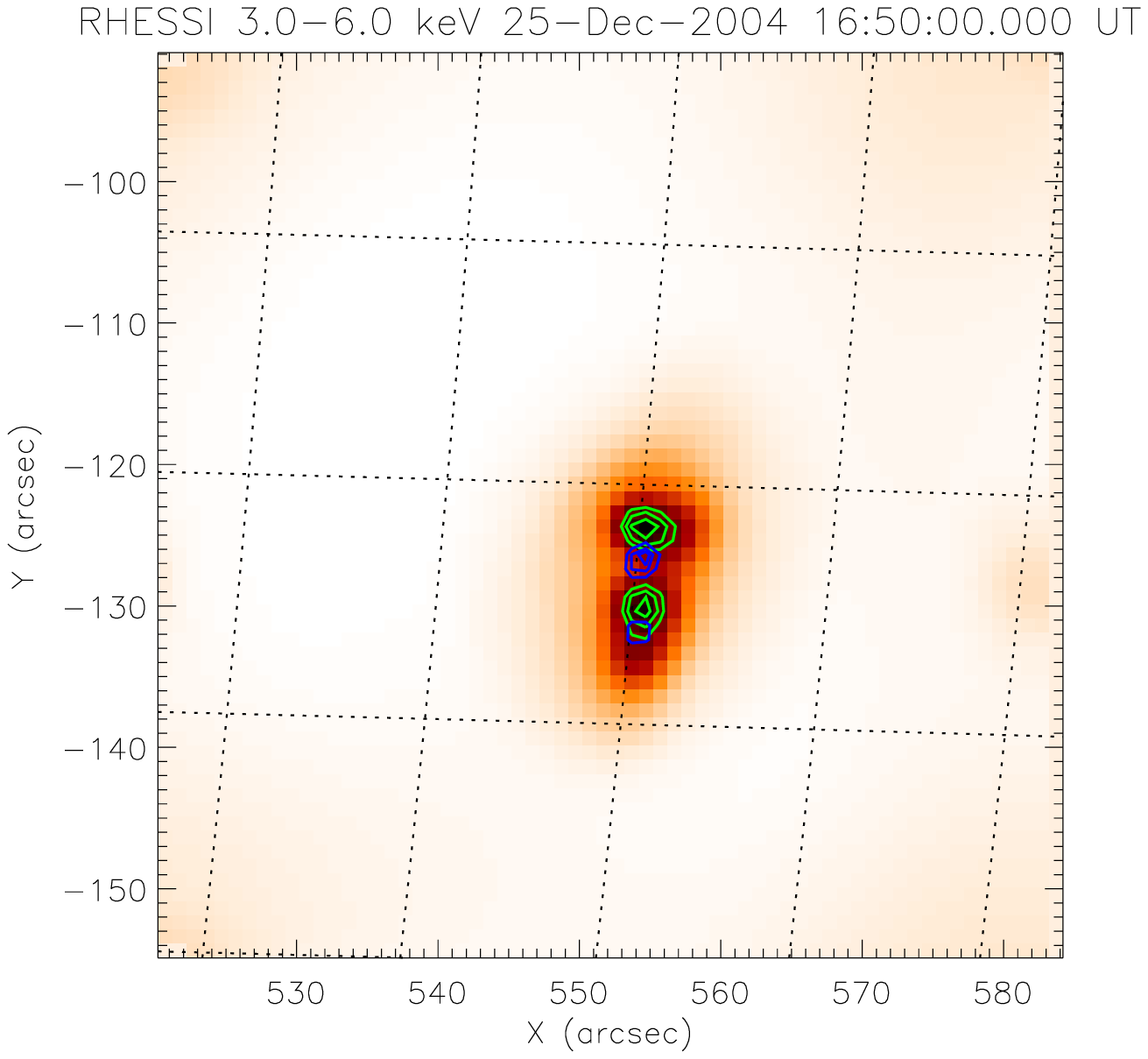}} &
  \hspace*{-1.4cm}
  \subfloat[]{\includegraphics[scale=0.5]{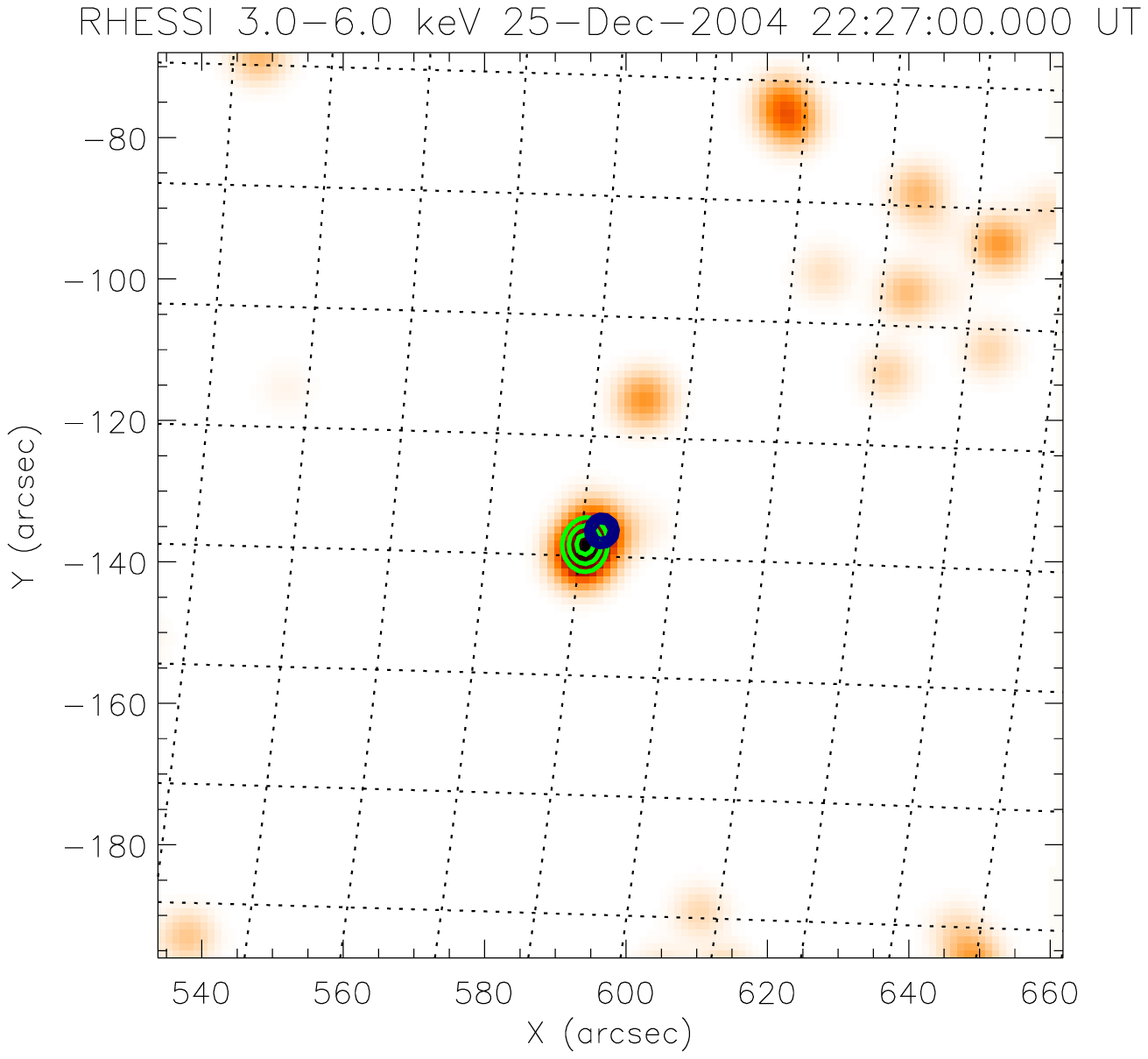}} \\
  \end{tabular}
  \caption{HXR images using the Pixon algorithm in the OSPEX/RHESSI routine for the six events in our final shortlist. The green contours show 70\%, 80\%, and 90\%
  emission levels in the 3-6 keV band. The blue contours show the 70\%, 80\% and 90\% emission levels in the 6-9 keV band.}
  \label{fig:pixon}
\end{figure*}
%Our goal is to estimate the energetics involved in producing the nonthermal electron population that we
%To estimate the energetics of the events in the final shortlist, we devised a two prong approach. Most of the energy released in the form of non-potential
%magnetic energy from the reconnection is transferred into heating plasma and accelrating particles. The heated plasma undergo further collisons and dump
%their energy back in the form of SXR radiation and heat. Most of the accelerated particles from the reconnection site rains down to denser chromosphere, where 
%they dump their energy through collisons and produce radiation and heat. However some of these accelerated particles escape into interplanetary space through
%open magnetic field lines. These are the particles detected by ACE/EPAM. To understand the energetics it is important to study the energy carried by the escaping
%electrons as well as the main non-thermal population at the reconnection site.
%We have divided this section into two parts for analyzing both the population of electrons.

\subsection{Escaping electrons detected at 1AU}
Electrons arriving at 1 AU are recorded in four energy channels between 38 and 315 keV by ACE/EPAM. We use ACE/EPAM 12 second data
\citep{gold1998}. The peak mean energies of the four
channels, DE1-4 are 45,74,134 and 235 keV respectively. %Pre-event substracted peak spectrum of the all the events are calcualted and tabulated in Table 2. 
Based on the flux at the peak of the event recorded at the mean energy of each ACE/EPAM energy channel, the differential energy spectrum of the particles 
can be represented as,
\begin{equation}
\label{eq2}
  dJ/dE \propto E^{\delta_{1}} 
   \end{equation}

 %  \begin{figure*}
%\begin{tabular}{cc}
  %\subfloat[The top panel shows RHESSI lightcurves for the Dec 25 2004 event from 3-100 keV.
    %This is a rather weak event, owing to which the photon counts above 50 keV are low. The bottom panel shows GOES X ray flux for the same event. 
   % It is clear that the SXR activity is very low; there is only a very small flux enhancement around 22:28 UT. This time is labeled ``Onset at GOES'' in
   % Table 1.] {\includegraphics[width=\columnwidth]{rhessigoes.png} } &
  %\subfloat[Top panel shows the ACE/EPAM 5 min averaged data showing the rise and decay profile of the event on Dec 25 2004, with $t_{sun}=$ 22.30 UT.
   % The bottom panel shows the WIND/WAVES RAD1 and RAD2 receiver voltages covering 20-14000 Hz. It is evident from the bottom panel that the IP electron 
   % signature (labeled ``Onset at WAVES'' in Table 1) is within a minute of $t_{sun}$.]{\includegraphics[width=\columnwidth]{acewind.png}} \\

  %\end{tabular}
 % \caption{Observations for the second event on Dec 25 2004. }
%\end{figure*}

\begin{figure*}
\begin{tabular}{ccc}
\hspace*{-1.3cm} 
  \subfloat[]{\includegraphics[scale=0.65]{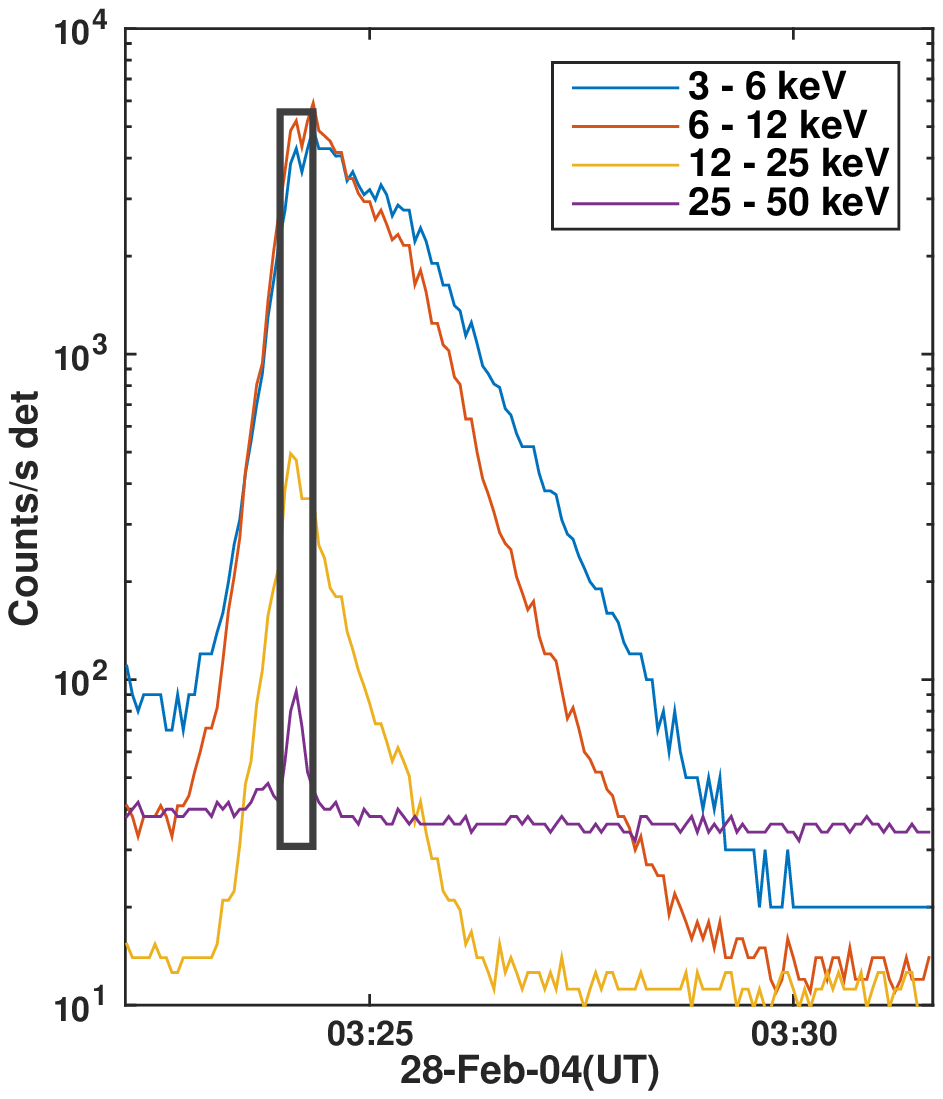}} &
  \hspace*{-1cm} 
  \subfloat[]{\includegraphics[scale=0.65]{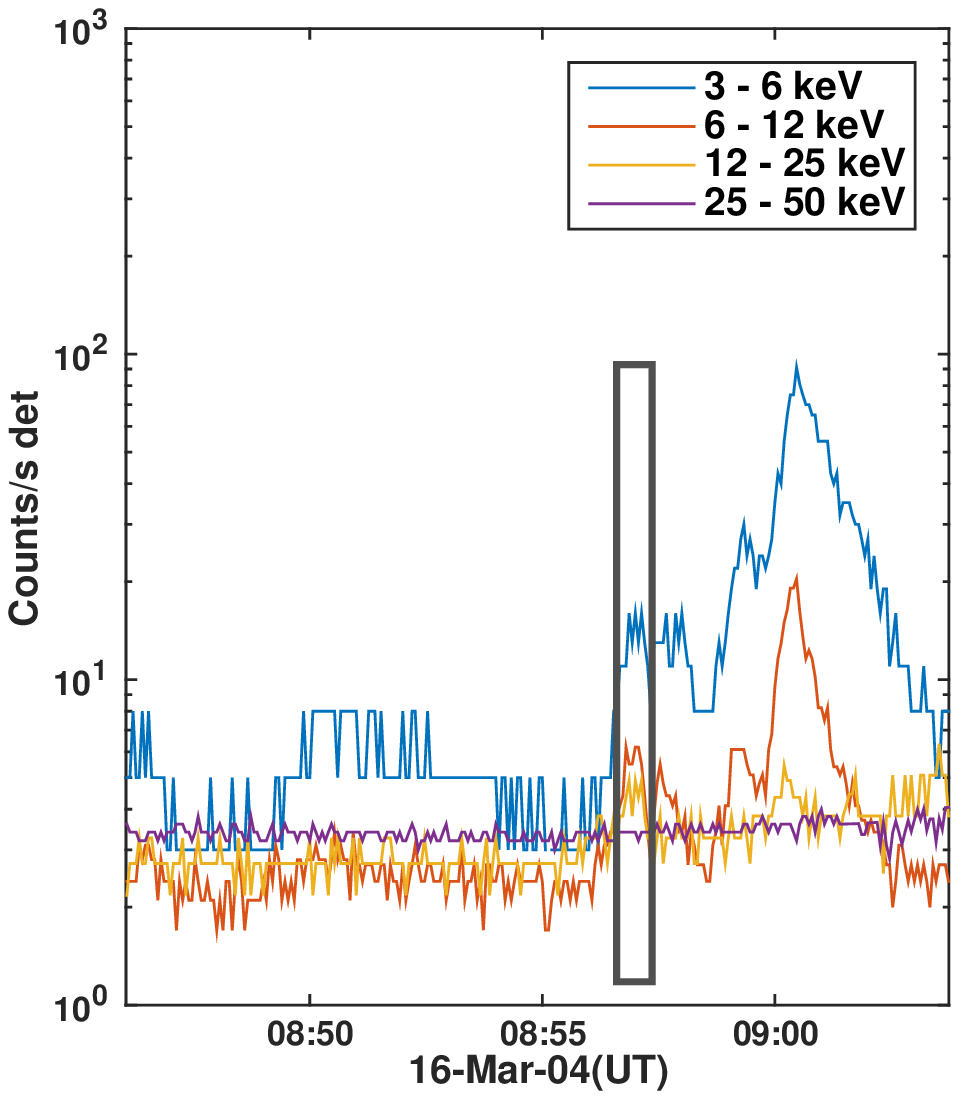}} &
  \hspace*{-1cm} 
  \subfloat[]{\includegraphics[scale=0.65]{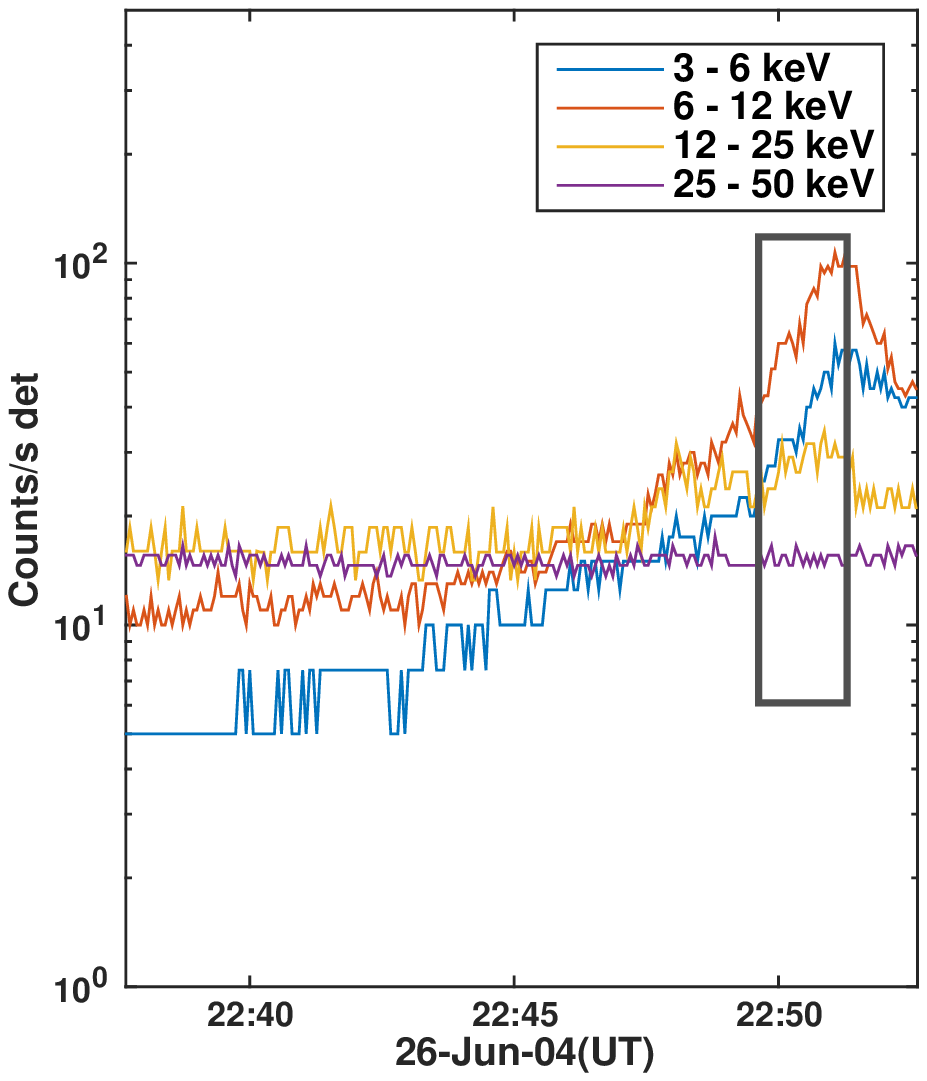}} \\
  \hspace*{-1.3cm} 
  \subfloat[]{\includegraphics[scale=0.65]{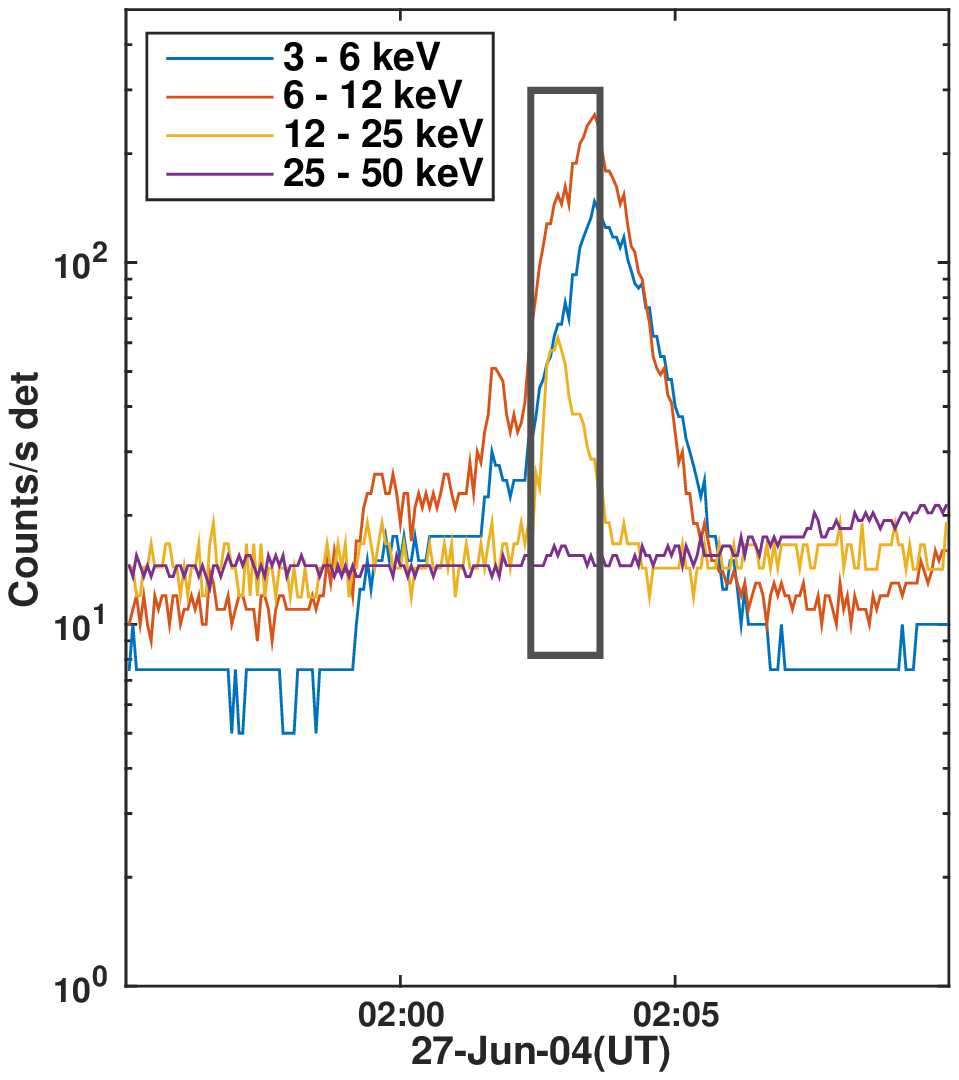}} &  
  \hspace*{-1cm} 
  \subfloat[]{\includegraphics[scale=0.65]{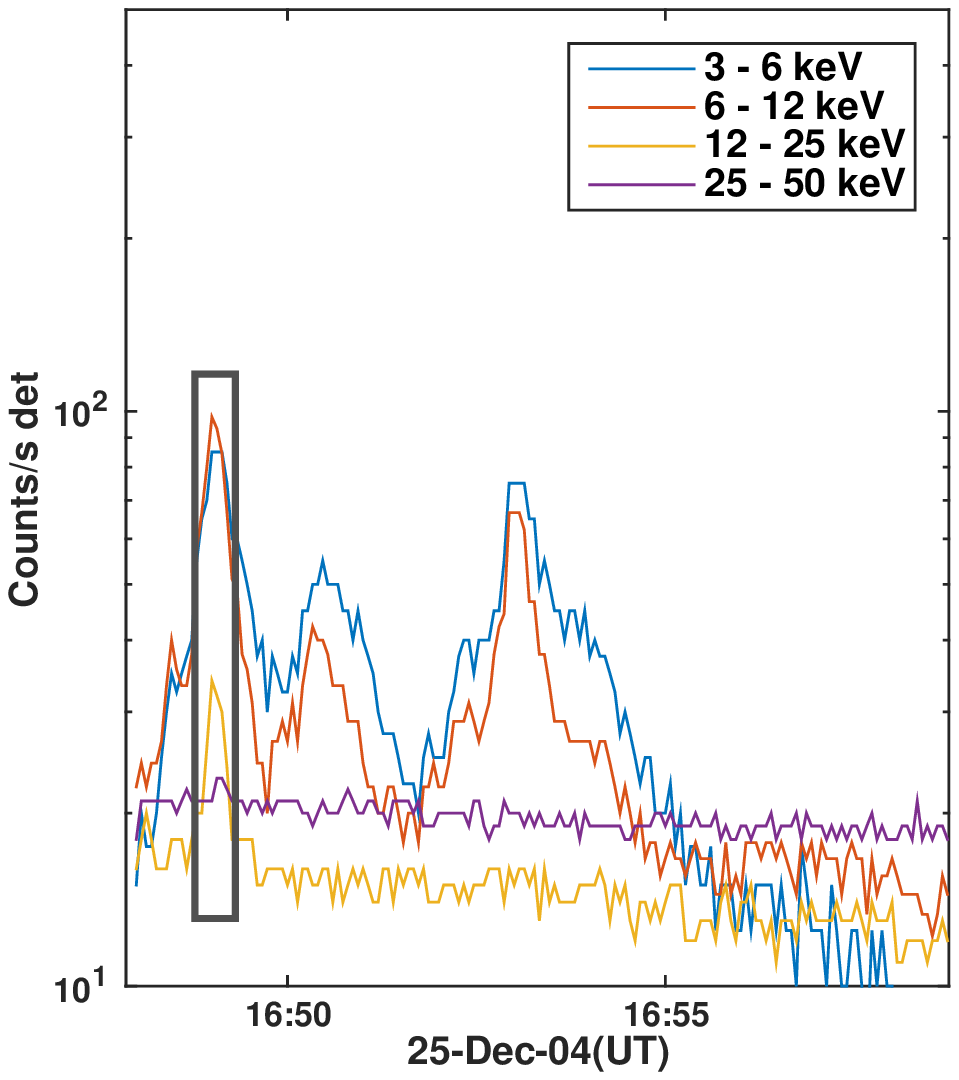}} &
  \hspace*{-1cm} 
  \subfloat[]{\includegraphics[scale=0.65]{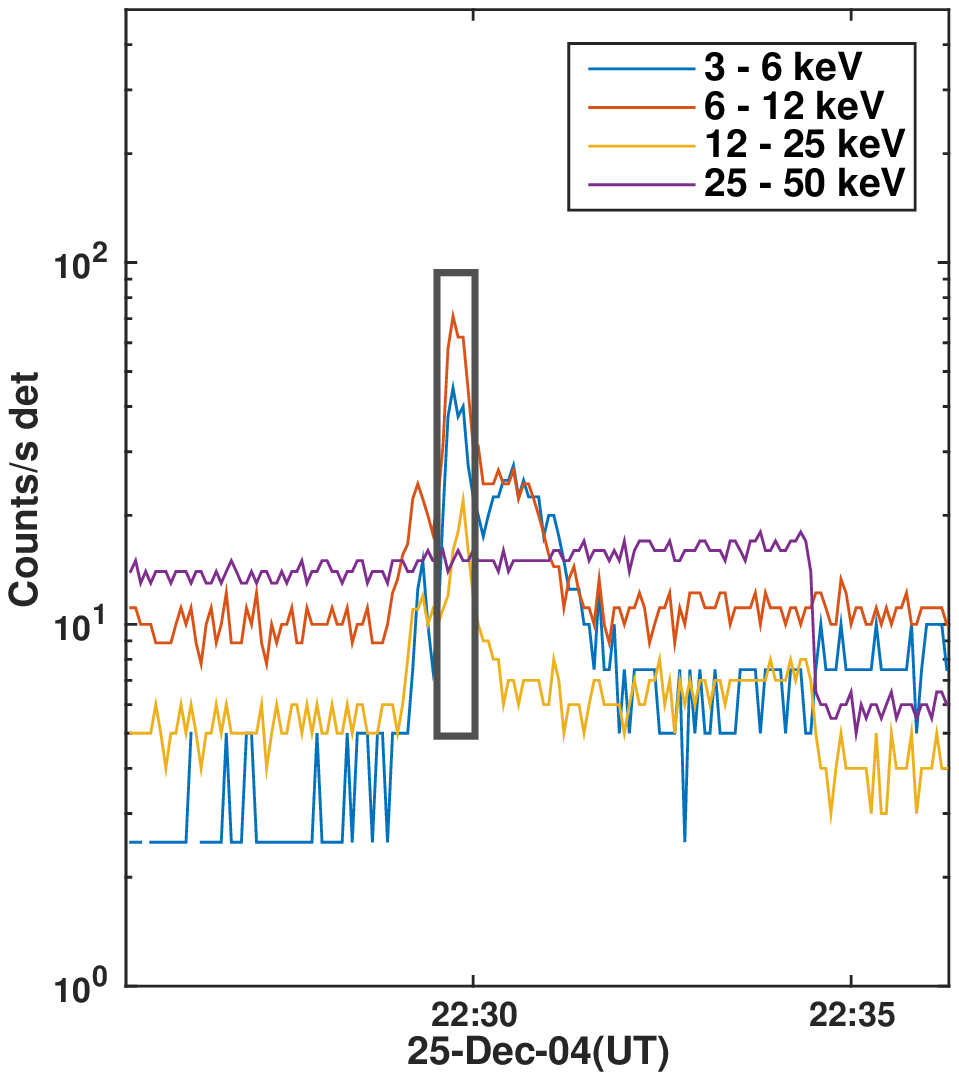}} \\
  \end{tabular}
  \caption{The HXR lightcurves for each of the six shortlisted events. The energy bands used for plotting the graph were 3-6 keV,6-12 keV, 12-25 keV, 25-50 keV, 50-100 keV.
           Due to weak nature of the events, there is little data in the highest energy band for most events. The shaded recatangle show the time strips used to generate
           the photon spectrum used in OSPEX.}
   \label{fig:ligtcurve}
\end{figure*}
\begin{table}
\caption{Final shortlist and spectral parameters}
\label{Q-sun events}
\begin{tabular}{cccccc}     % define the column alignment
                           % l: left, c: center, r: right
  \hline                   % horizontal line
 Date            &RHESSI Peak  & Flare Position   & $\gamma$                &     $\delta_{1AU}$     & $\delta_{hxr}$                 \\
                 &Time(UT)         & Heliocentric   &                         &                         &            \\
                 &             & x,y(arcsec)          &                         &                          &              \\
  \hline
Feb 28,2004	 &03.24	       & 697, 301   	     &  3.69  		      &    4.14               &4.89    \\
Mar 16,2004      &08:56        & 841, -144             &  4.38                  &    5.25               &5.54            \\                                                                                                                  
June 26,2004     &22.50        & 943, -162            &  4.66                  &    5.45               &5.80                         \\
June 27,2004     &02.03        & 931, -176             &  4.24                  &    5.10               &5.13        \\
Dec  25,2004     &16.49        & 552, -123             &  3.25                  &    4.39               &4.43         \\
Dec  25,2004     &22.29        & 598, -132             &  3.10                  &    4.12               &4.06                 \\

\hline
\end{tabular}
\label{tab:spectrum}
\end{table}
\noindent

\begin{figure*}
\begin{tabular}{ccc}
  \hspace*{-1.7cm}
  \subfloat[]{\includegraphics[scale=0.60]{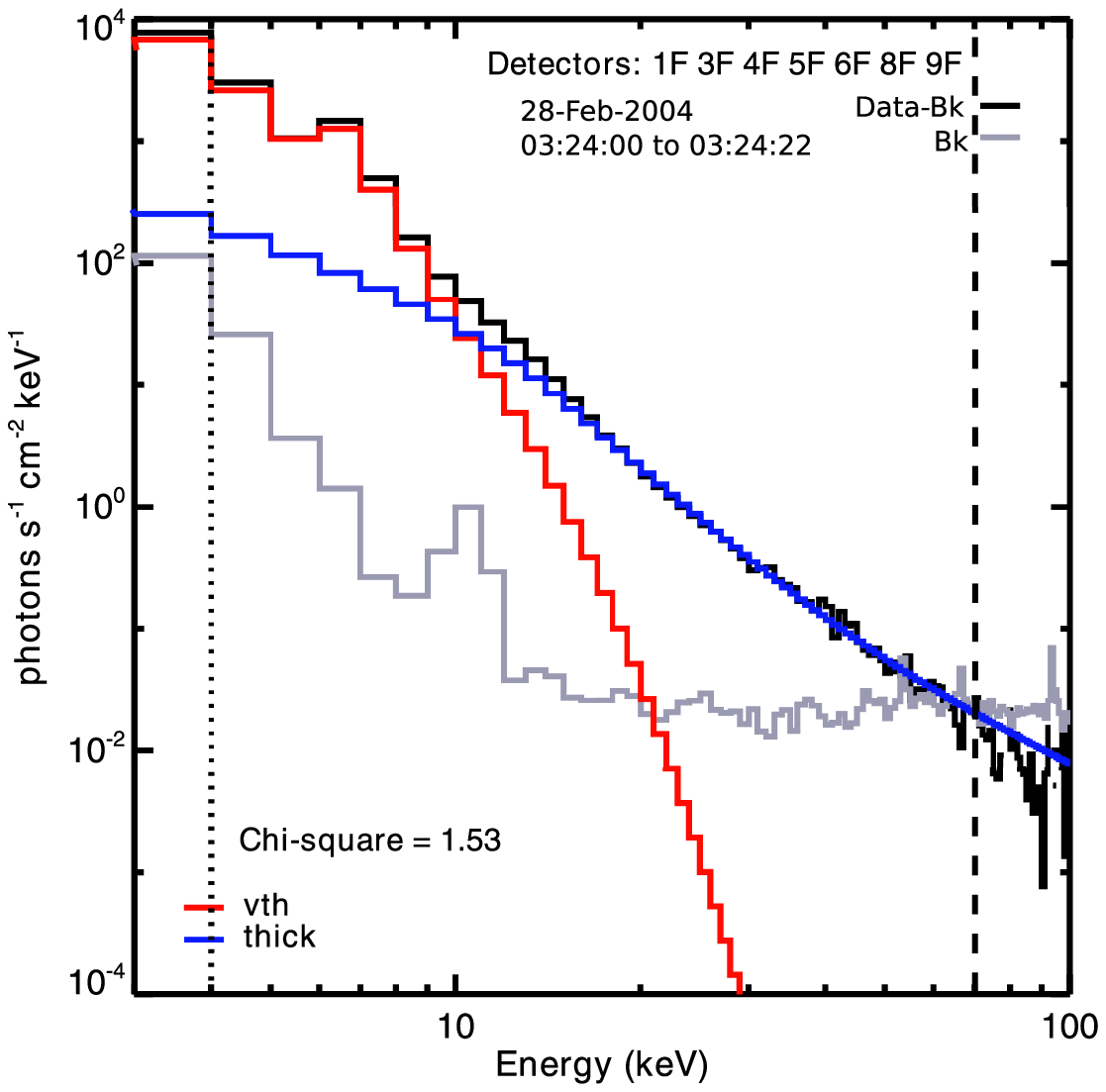}} &
  \hspace*{-0.6cm}
  \subfloat[]{\includegraphics[scale=0.65]{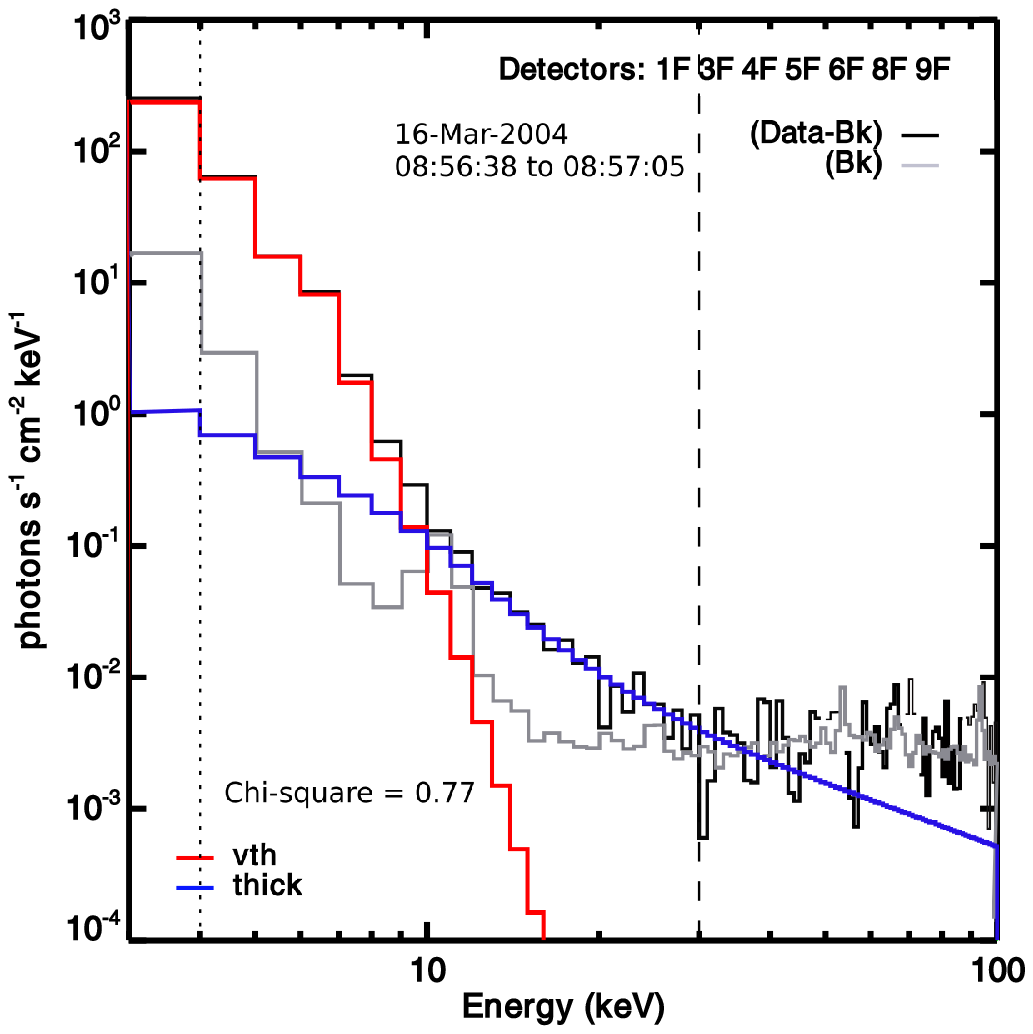}} &
  \hspace*{-0.6cm}
  \subfloat[]{\includegraphics[scale=0.63]{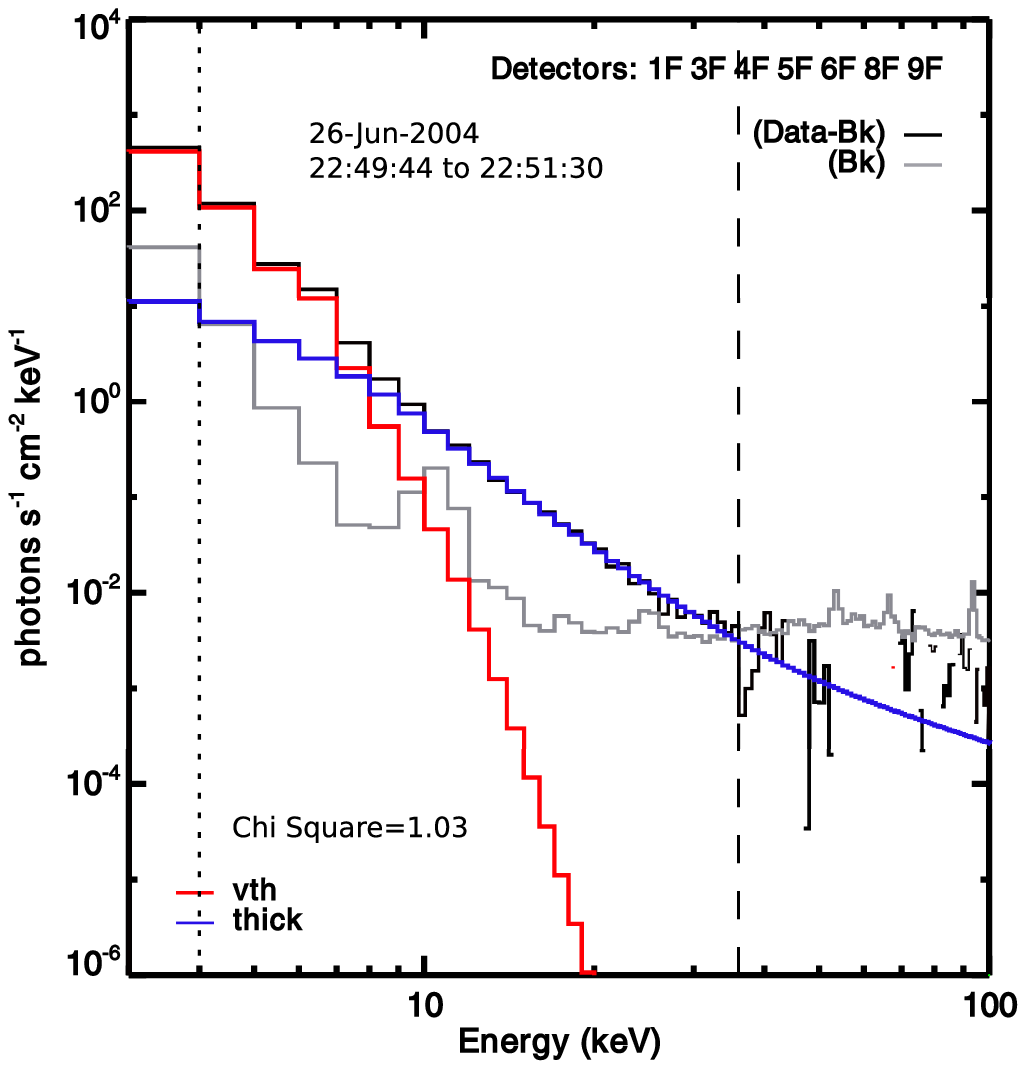}} \\
  \hspace*{-1.7cm}
  \subfloat[]{\includegraphics[scale=0.65]{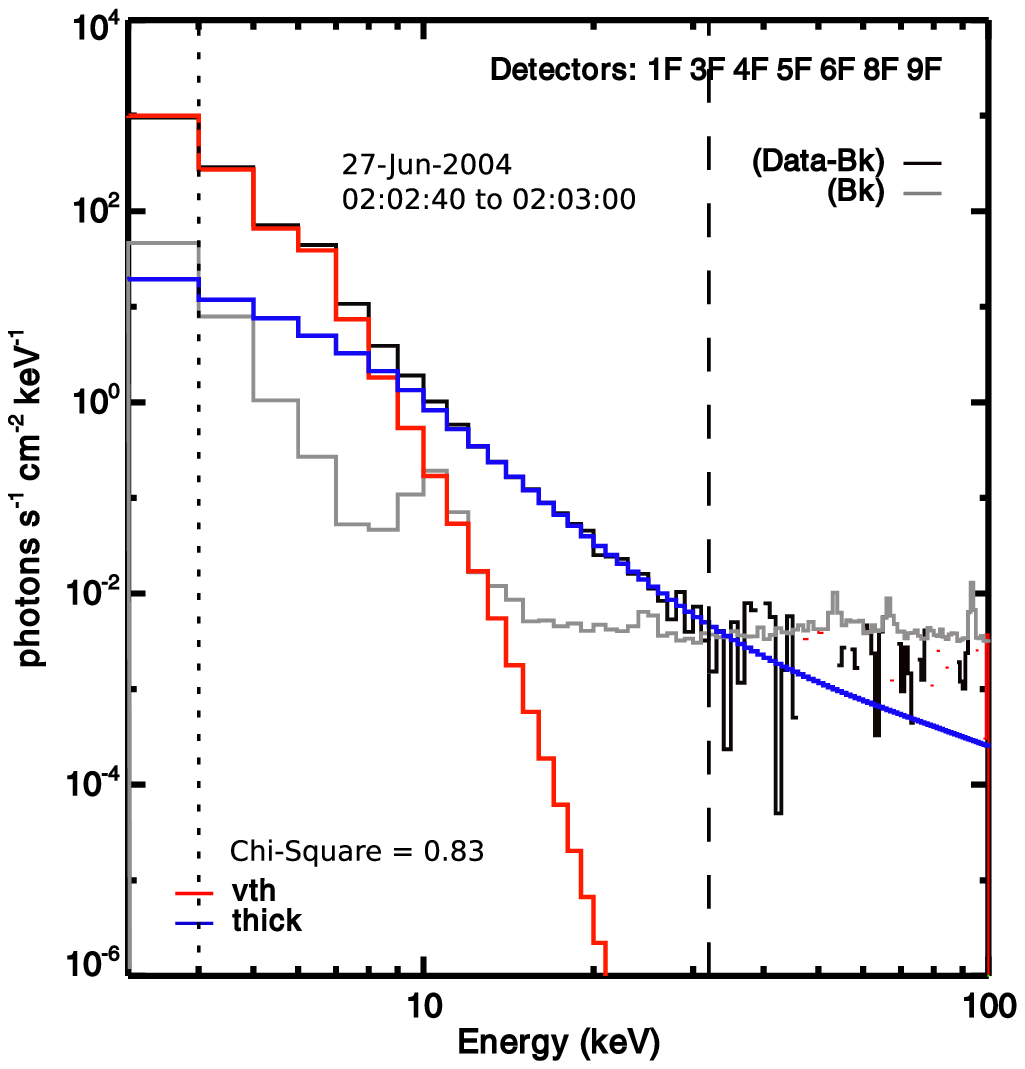}} &
  \hspace*{-0.6cm}
  \subfloat[]{\includegraphics[scale=0.65]{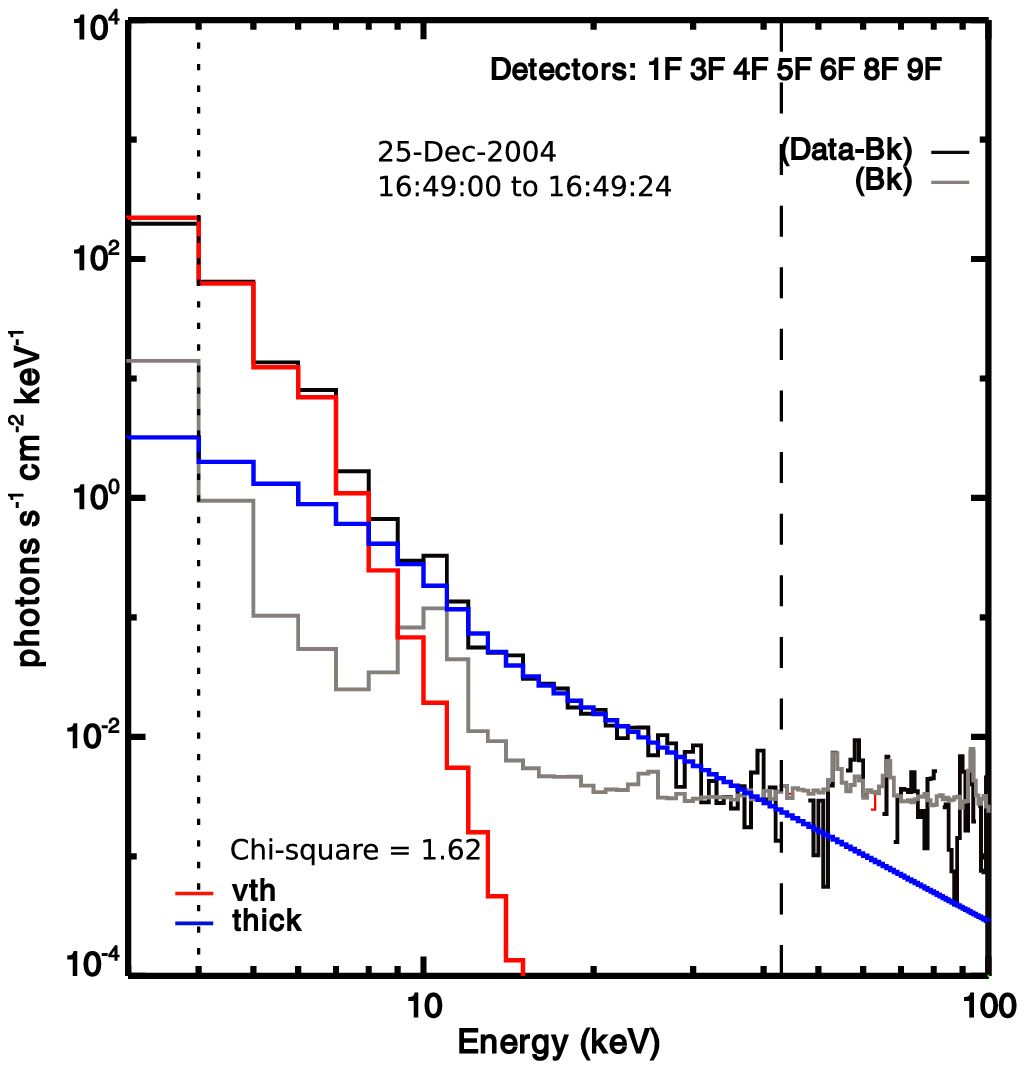}} &
  \hspace*{-0.6cm}
  \subfloat[]{\includegraphics[scale=0.65]{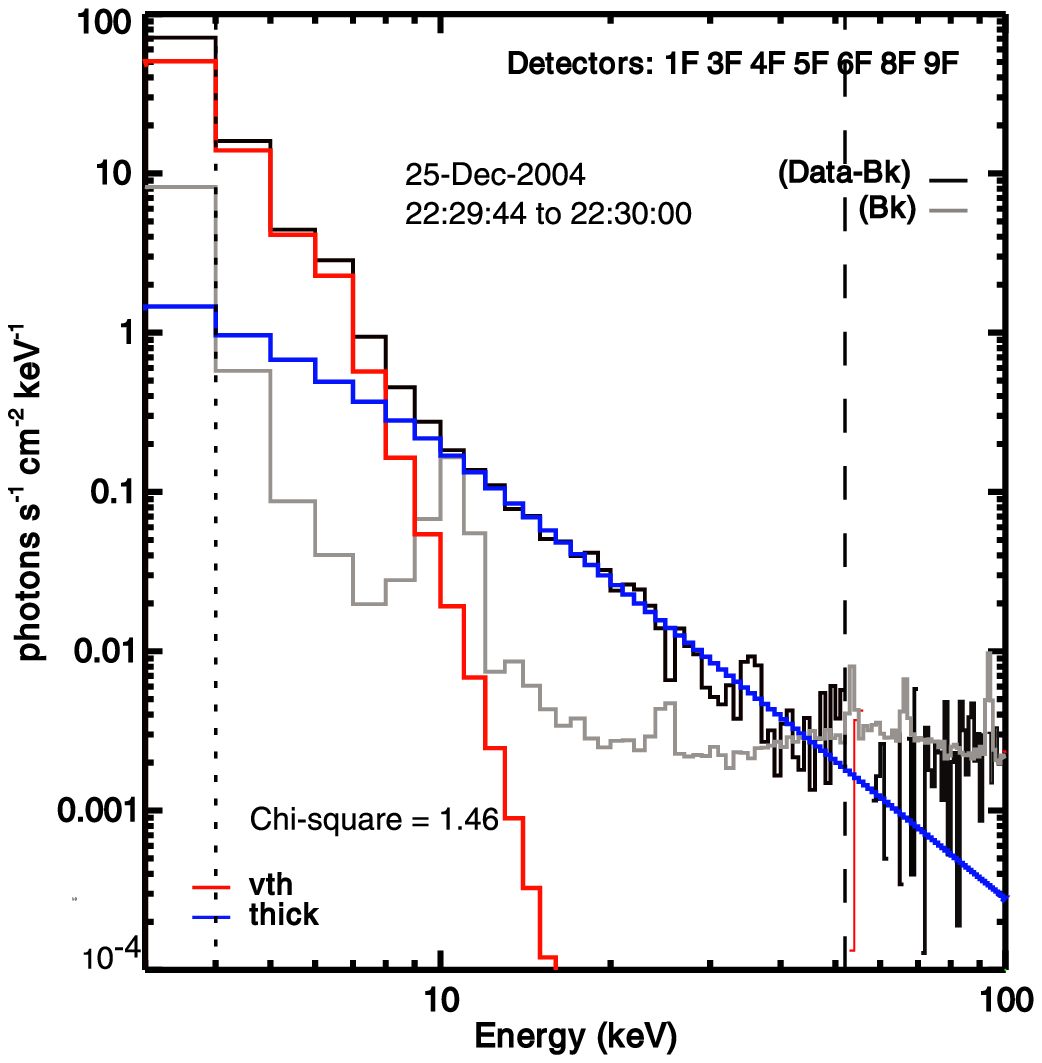}} \\
  \end{tabular}
  \caption{The OSPEX spectral fits for each of the six shortlisted events. RHESSI front detectors 1,3,4,5,6,8,9 were used to accumulate the data. Since the events are weak, we restrict the fit to energy ranges which have
   counts above the background emission levels. The dashed vertical lines indicate the range of energies used for the fit.
   A photon spectral index $\gamma$ is obtained through
   a broken power law fit (not shown). The emission measure and thermal temperature are obtained from variable thermal function fit.
    The thick target fit gives 
   the hxr producing electron spectral indice $\delta_{hxr}$. Results are tabulated in Table \ref{tab:spectrum}. }
   \label{fig:thick}
\end{figure*}
    
where $\delta_{1}$ is the spectral index of electrons escaping from the coronal acceleration site and detected in-situ at 1 AU by ACE/EPAM. This is distinct from 
the spectral index $\delta_{\rm hxr}$ of the HXR producing electrons, which is estimated in \S~3.2. The values of these indices for 
each event are summarized in Table \ref{tab:spectrum}. For a given event, one would expect that $\delta_{1}$ and $\delta_{\rm hxr}$ be similar to each other, if the electrons propagate in a reasonably scatter-free manner from the Sun to the Earth.

The fit to the spectra of electrons detected in-situ at 1 AU by ACE/EPAM for all the events are shown in Fig \ref{fig:delta}. The fits are constructed using the peak count in each energy bin.
All the events typically show a break in the spectrum at around 74 keV (Figure \ref{fig:delta}). By contrast, the break energy for larger flares tends to 
be around 
48 keV (e.g., \citet{krucker2007}). 
The break energy in the spectrum presumably
indicates the energy above which wave-particle interactions are unimportant \citep{kontar2009}. %to this break in the spectrum. 
%However, the break-energy is dependent upon the type and intensity of events. 
We note that \citet{krucker2007} used WIND/3DP data for their analysis. Of the events in our final shortlist,
four have reliable WIND/3DP data. The break energy for these events as ascertained from WIND/3DP data is $\approx$ 66 keV, which corresponds to 
the mean energy of the corresponding channel on that instrument. 
%This is expected because the mean energy of the DE2 channel on
%WIND/3DP is 66 keV.
%This observed increase in break energy matches well with simulations (Kontar 2009, Fig 2). (explain reason)

 \subsubsection{Number flux, power and energy in Escaping electrons}
Having fitted the differential electron flux at 1 AU $F(E)$ (${\rm electrons\, s^{-1}\,sr^{-1}\,keV^{-1}\,cm^{-2}}$) to a power law of the form
$F(E) = F_{0} (E_{0}/E)^{\delta_{1}}$, (Table \ref{tab:spectrum}) it follows that the flux of electrons above $E_{0}$ is
\begin{equation}
\int_{E_{0}}^{\infty} F(E) dE =  \frac{F_{0} E_{0}}{(\delta_{1} - 1)} \, \, \, {\rm electrons\, s^{-1}\,sr^{-1}\,cm^{-2}} \, .
\label{eqfl1}
\end{equation}
The quantity $E_{0}$ is the break energy in ACE/EPAM data (typically around 74 keV)  above which the power law fit is valid 
and $F_{0}$ is the peak differential energy flux  at $E_{0}$. The number of escaping electrons per second above
$E_{0}$ detected by ACE/EPAM at 1 AU and the power carried by them are given by
{\begin{eqnarray}
\label{eqfl2}
\nonumber
\Omega (1 AU)^{2} \int_{E_{0}}^{\infty} F(E) dE = \Omega (1 AU)^{2}   \frac{F_{0} E_{0}}{(\delta_{1} - 1)} \, \, \, {\rm s^{-1}} \, , \, \, \, {\rm and}\\
\noindent
\Omega (1 AU)^{2}  \int_{E_{0}}^{\infty} E F(E) dE =  \Omega (1 AU)^{2}   \frac{F_{0} E_{0}^{2}}{(\delta_{1} - 2)} \, \, \, \, {\rm erg\, s^{-1}}
\end{eqnarray}}
respectively, where $\Omega$ is the solid angle spread of the electrons at 1 AU, which we take to correspond to a
cone of 30$^{\circ}$ \citep{lin1974,krucker2007}. The factor of $(1 AU)^{2}$ arises from assuming that the electrons are spread 
uniformly over a sphere of radius 1 AU. Multiplying the expression for power by the duration of the impulsive electron event at 1 AU yields
the energy carried by the escaping electrons. The power (${\rm erg\,s^{-1}}$) and energy (ergs) for each of our shortlisted electron spikes detected at 
1 AU are listed in Table \ref{tab:ennumbers}.

%The ACE/EPAM electron channels won't detect electrons below 27 KeV. 

%{\color{blue} STOPPED HERE, PM AUG 26 2016}

\subsection{HXR producing electrons}
The RHESSI HXR photon spectrum for each event was analysed using SSW/OSPEX. The assumed HXR photon spectrum consists of a thermal core and a long non-thermal
tail. 
%The time resolution of the spectrum is 4s corresponding to one spin cycle. 
%We observe that the duration of the HXR event decreases as we progress 
%higher in energy (Figure \ref{fig:ligtcurve}). 
The event duration in the lowest available energy channel (which has a reasonable number of counts) was used 
as integration time for computing the spectra.
We applied a broken power law fit to the non-thermal tail (Figure \ref{fig:thick}), 
with a variable thermal function for the
thermal core at low energies. The background spectrum dominates the event spectrum
above 35 keV for most of the events. We perform these
fits only for energies with photon counts above the background. 
The photon spectral indices $\gamma$ are listed in 
Table \ref{tab:spectrum}. 
We note that break energies for large flares are $\sim$ 45 keV \citep{krucker2007,oka2013}. 
%Hence we restricted the fit to the energy above which the spectra got contaminated with background photons.
The fit for all the events are shown in Figure \ref{fig:thick} . The HXR  spectral indices for all the events are shown 
in Table \ref{tab:spectrum} . 
%A very strong
%correlation with coefficeint 0.90 is observed between the photon spectral index
%$\gamma$ and the in-situ electron spectral index $\delta_{esc}$. 
Assuming the injected electron spectrum in the coronal acceleration site is
same as the interplanetary electron spectrum, \citet{datlowe1973}  noted that a thick target
model for HXR emission gives $\delta_{1} = \gamma +1$ while a thin target one gives $\delta_{1} = \gamma - 1$. 
 
 We find that the photon spectral index $\gamma$ is related to the electron spectral index $\delta_{1}$ observed
at 1 AU via $\delta_{1} = \gamma +1$ with a correlation coefficient of 0.86 (Figure \ref{fig:kruckermiddle}a). Our finding that $\delta_{1} = \gamma +1$ thus suggests that the HXR emission is produced via the thick target
bremsstrahlung process.
%The spectral indices of electrons detected near 1AU
%from ACE/EPAM data and the photon spectral index from RHESSI were fitted with a linear model to investigate any spectral dependence. We found a model of 
%the form $\delta_{esc}$ = $\gamma$ +0.90, fits with data with high degree of confidence. 
%This strongly suggests that most of the electrons lost their energy
%through collisons or a thick target bremstrahlung emission is at work.

\begin{figure}%

    \subfloat[Delta indices]{{\includegraphics[scale=0.7]{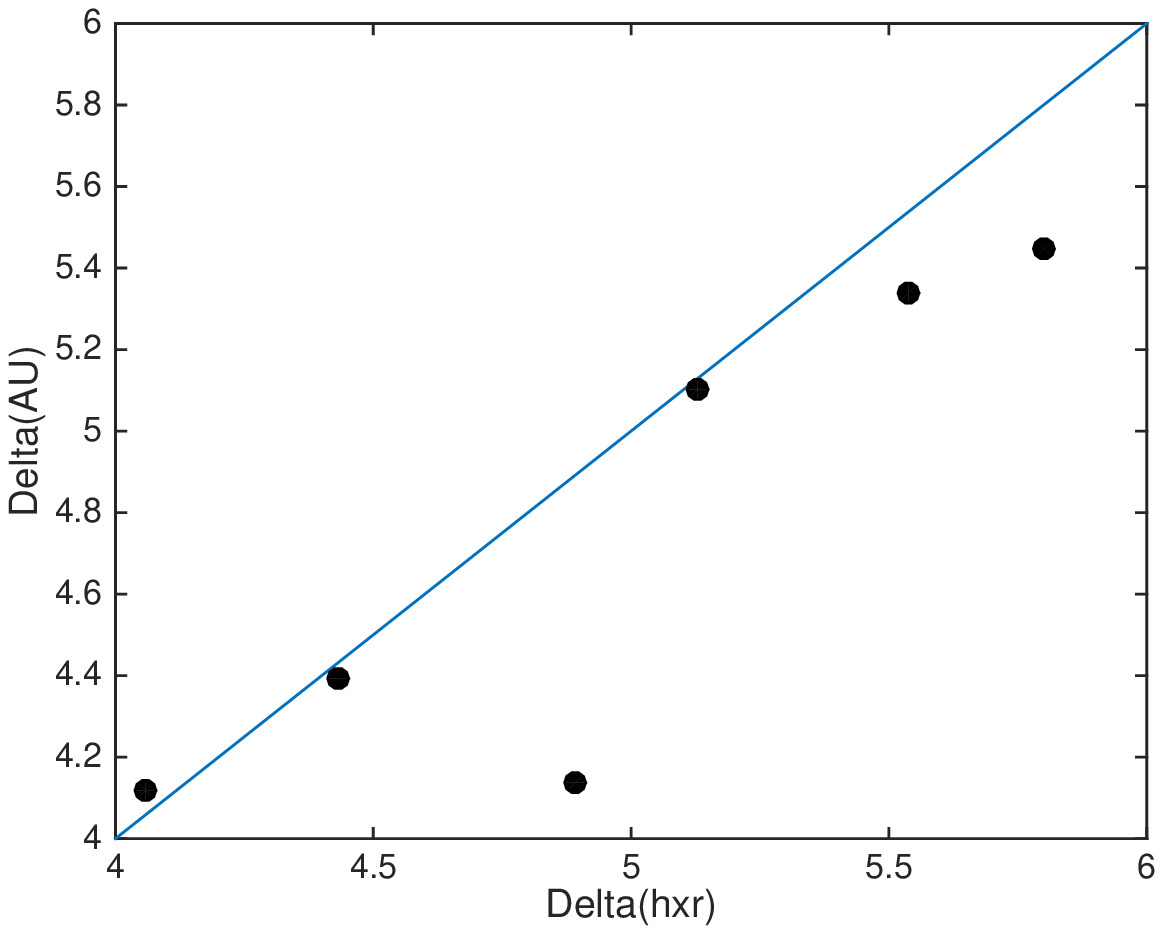} }}%
    \caption{A scatterplot between $\delta_{1AU}$ and $\delta_{hxr} $. The two indices are linearly related with a correlation coefficient of 0.91}%
    \label{fig:deltavsdel}%
\end{figure}

We therefore employ a thick target model, with a isothermal function for
the thermal component. We require that the thermal temperature and emission measure of the variable thermal function remain constant from the previous fit for
estimating the photon spectral index.

The thick target function gives the integrated electron flux at the HXR production site, 
injected electron spectrum and low energy cutoffs. It also inverts the photon spectrum to give a broken power law electron spectrum, with the break occurring at the injection energy. 
The spectral index below the injection energy is denoted by $\delta_{hxr}$, and is tabulated in Table \ref{tab:spectrum}. Figure \ref{fig:deltavsdel} shows a scatter plot of $\delta_{hxr}$ and $\delta_{1}$.
The results of our analysis in Sec 3.1 reveal that the spectral index of the HXR producing electrons and that of the electrons detected
at 1 AU are not very different. They exhibit a linear correlation with a coefficient of 0.91. This implies that the electron population emitting HXR radiation and the one detected at 1 AU have a common origin in the coronal acceleration region. 
\begin{table}
\caption{Number of HXR producing and escaping Electrons
}
\begin{tabular}{cccc}     % define the column alignment
                           % l: left, c: center, r: right
  \hline                   % horizontal line
 Date            &Number of                  & Number of              & Escaping fraction                          \\
                 &hxr producing              &  escaping               &  $\eta= (N_{esc}/N_{hxr})\% $                          \\
                 &electrons$N_{hxr}$           & electrons$N_{esc}$                                                            \\
                 &(>74 keV)                    &(>74 keV)                                                              \\
 \hline
Feb 28,2004	 &1.82$\times10^{33}$	       &1.16$\times10^{32}$           &  6.36  		          \\
Mar 16,2004      &3.60$\times10^{30}$       & 5.35$\times10^{30}$             &  148.3                             \\                                                                                                                  
June 26,2004     &2.31$\times10^{31}$        & 1.07$\times10^{30}$             &  4.64                                       \\
June 27,2004     &5.21$\times10^{31}$        &6.03$\times10^{31}$             &  115.7                     \\
Dec  25,2004     &1.89$\times10^{32}$       & 3.31$\times10^{31}$             &  17.44                        \\
Dec  25,2004     &8.57$\times10^{31}$        & 3.65$\times10^{31}$            &  42.62                            \\

\hline
\end{tabular}
\label{tab:numbers}
\end{table}

 %{\color{blue} TOMIN - STOP HERE}
 \begin{table*}
\caption{Fit Parameters of the Events
}
\begin{tabular}{cccccc}     % define the column alignment
                           % l: left, c: center, r: right
  \hline                   % horizontal line
 Date            &Thermal                 & Emission                         & Energy               &  Density of                              & Volume                \\
                 &Temperature$(MK)$        & Measure$(10^{46} cm^{-3})$        & Cutoff $(keV)$         &  thermal plasma$(10^{10} cm^{-3})$         &  ($10^{25} cm^{3}$)        \\
  \hline
Feb 28,2004	 &13.53	       & 9.27   	     &  14.4 		      &  4.50                 &5.05    \\
Mar 16,2004      &11.08        & 0.95             &  12.34                  &    3.33               & 0.86            \\                                                                                                                  
June 26,2004     &11.32        & 1.08             &  9.38                  &     2.10              &2.43                        \\
June 27,2004     &11.50        & 1.73             &  10.01                  &    2.34               &3.15       \\
Dec  25,2004     &10.05        & 0.68            &   11.2                 &    2.22               &1.39         \\
Dec  25,2004     &10.23        & 0.13            &  10.89                 &    0.84             &1.78                \\

\hline
\end{tabular}
\label{tab:fitparas}
\end{table*}

\subsubsection{Number flux, power and energy in HXR producing electrons}

The number of electrons per second involved in producing the HXR emission ($\dot{N}_{>E_{c}}$) is computed for each event by
the SSW/OSPEX thick target emission model. This procedure also yields the cutoff energy $E_{c}$ 
 and the power law index $\delta_{\rm HXR}$ of the electron spectrum above  $E_{c}$. Since the number of electrons escaped and detected at 1 AU,
 were estimated only above 74 keV, we scale the corresponding number of HXR electrons above 74 keV using equation Eq~\ref{eqscale}.
 \begin{equation}
  \label{eqscale}
  \dot{N}_{>74 keV}=\dot{N}_{>E_{c}}\bigg(\frac{E_{c}}{74}\bigg)^{\delta_{hxr}-1}
 \end{equation}

\begin{figure}%
    \centering
     \subfloat[Thick target relation of spectral indices]{\includegraphics[scale=0.55]{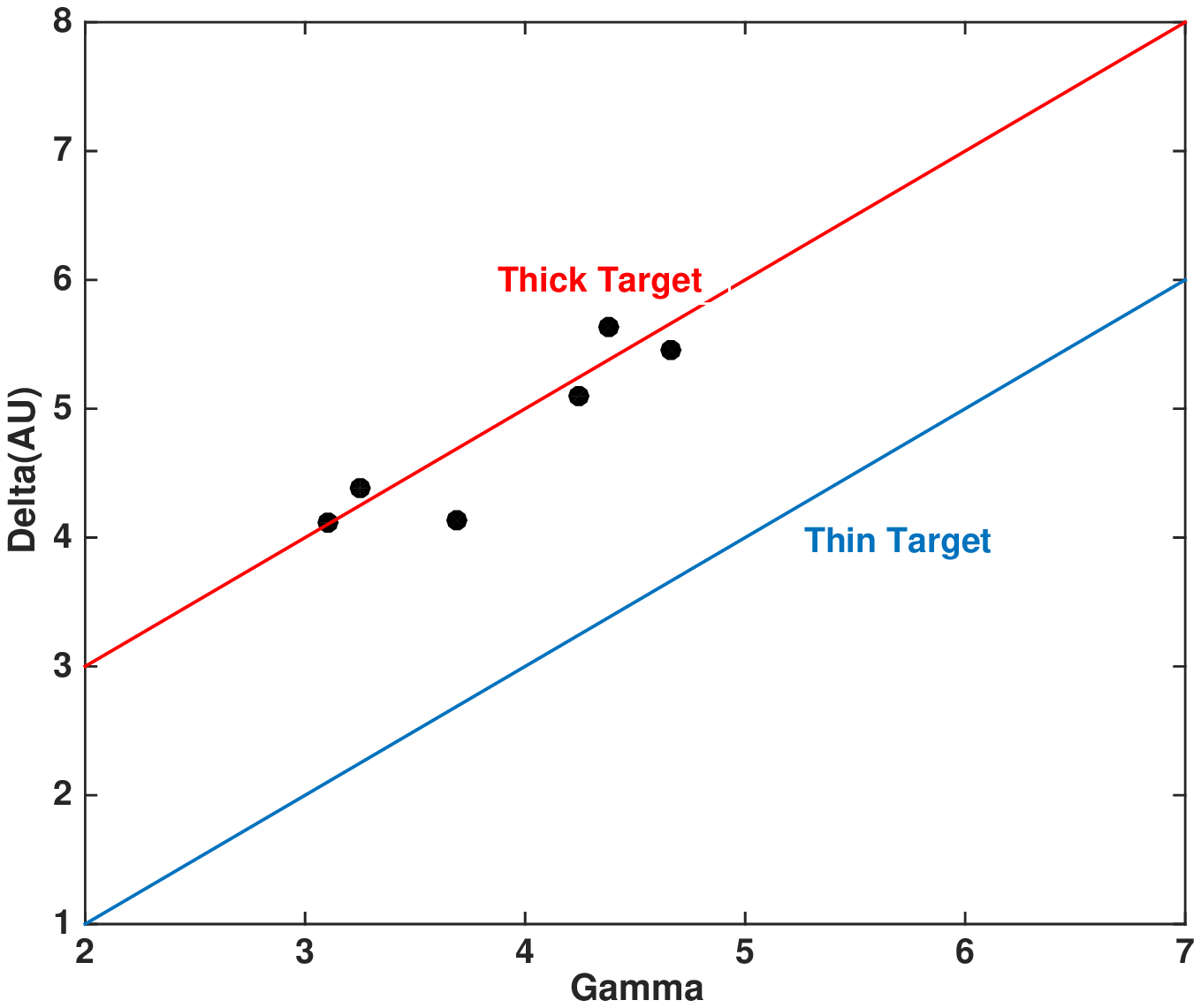}} \\
    \subfloat[Spectral dependence of $\gamma$ and $\delta_{1}$]{\includegraphics[scale=0.25]{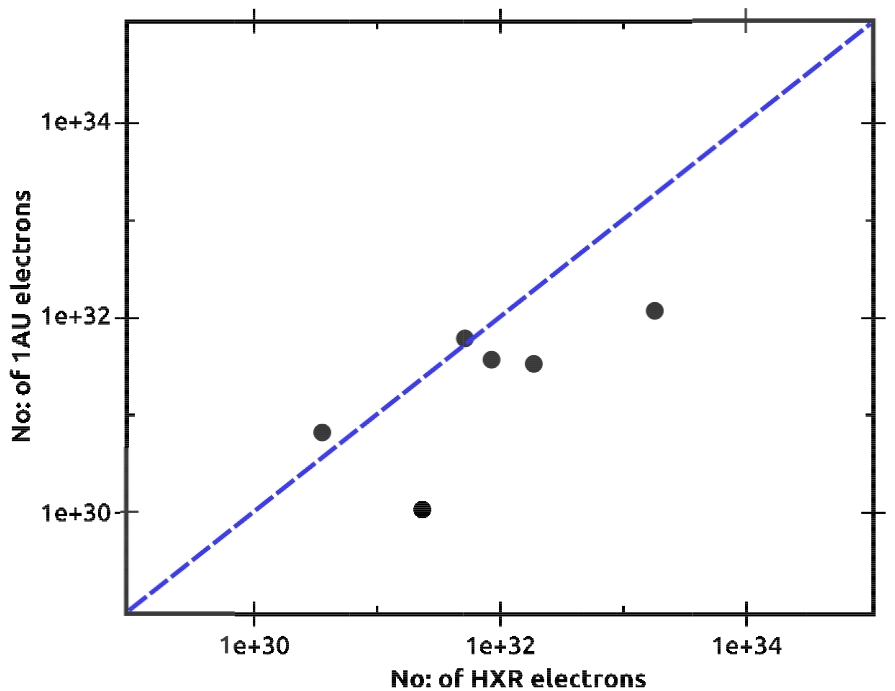}} 
  
    \caption{ 
    The top panel shows a scatterplot of the photon spectral
    index $\gamma$ plotted against $\delta_{1}$ (Table 2). We note that the data points fit $\delta_{1} = \gamma + 1$ with a correlation
    coefficient of 0.94. This indicates that the HXR radiation is most likely produced via thick target emission.
    The bottom panel shows a scatterplot of the total number of $> 74$ keV electrons detected at 1AU plotted against the number of 
    $> 74$ keV electrons involved in HXR emission. }%
    \label{fig:kruckermiddle}%
\end{figure}

 Following the same logic used in writing Eq~\ref{eqfl2},
 the power in the HXR emitting electrons above $E_{c}$ is given by
\begin{equation}
\label{eqfl3}
  P=(74 keV)\dot{N}_{>74 keV}\frac{\delta_{hxr} -1}{\delta_{hxr} -2}
 \end{equation}
 Multiplying this quantity with the duration of the HXR emission yields the total energy in the nonthermal HXR producing electrons. 
 These quantities are listed in Table \ref{tab:numbers} for each of our shortlisted events. The ratio ($\eta$) of the escaping electrons to the HXR producing electrons ranges from $\approx$ 6 \% to over 100 \%. By comparison, $\eta$ for the larger flares reported in
 \citet{krucker2007} is typically only 0.2 \%. %{\color{red} PS STOPPED HERE MARCH 13 2017}
 %Also we note two events showing ratio which is greater than 100, indicating more electrons escaped
 %towards earth than rain down to the flare footpoints. 
 Some of these results are graphically depicted in figure~\ref{fig:kruckermiddle}. Figure \ref{fig:kruckermiddle}a shows
 the plot of $\delta_{1AU}$, the energy spectral index of escaping electrons detected in-situ at 1AU and the $\gamma$, the photon spectral index. The spectral
 indices follow a functional relation closely resembling a thick target bremmstrahlung process. For big flares, this relation is not as pronounced
 \citet{krucker2007}. A plot of number of HXR producing electrons at the acceleration site and the number of escaping 
 electrons detected at 1AU is plotted in Figure \ref{fig:kruckermiddle}b. 
Comparing this graph with that for big flares (eg: Fig 3b, \citet{krucker2007}), we note that the fraction of escaping electrons for the weak events discussed in this paper is much larger. 

 %Here $\delta_{hxr}$ is the spectral index of hxr producing electrons derived from the thick target fit.$\dot{N}_{>E_{c}}$ is the 
%injection rate of electrons above cut-off energy $E_{c}$. This is the energy contained in the total non-thermal population, including the hxr 
%producing and the fraction escaping to interplanetary space. 

%{\color{red} NOT SURE IF WE NEED THE BIT ON THERMAL EMISSION}
%{\color{red} PS STOPPED HERE PM SEPT 4 2016.}

\subsubsection{Thermal HXR emission}

We next estimate the energy content of the thermal HXR emission generated by the accelerated electrons.
%We expect that a significant
%fraction of the energy contained in the non-thermal electrons to be converted into thermal HXR radiation when they collide with the dense chromosphere.
The thermal  energy $U_{th}$ is given by
\begin{equation}
 U_{th}= 3 k_{B}T \sqrt{\epsilon_{em} V}
\end{equation}
where $\epsilon_{em}$ is the emission measure and $V$ is the volume of the emission region. The emission measure
is a product of the electron density $n_{e}$ , ambient plasma density $n_{i}$ and the
volume $V$. The emission measure and temperature is determined using the thermal fit to the photon spectrum. The area A of the emission region was estimated by
first fitting a Guassian to the 70 \% flux contour levels of the RHESSI image
We used the Pixon algorithm which gives a sharper image as compared to the Clean algorithm \citep{brian2009}.The volume was
computed using the formula 
\begin{equation}
V=A^{3/2}
\end{equation}
Using this method we found the volume to be 
$ \sim 10^{25} {\rm cm}^{3}$ which is very low compared to volumes usually cited for radio noise storms or large flares \citep{prasad2006,krucker2007,lin1971}.
This value is similar to values for the smallest of microflares reported by \citet{Hannah_2008}.
For each event, we derive an upper limit
on the electron density using the emission measure, which we derived
by applying a thermal fit to the photon spectra. The emission measure was found to be around $10^{45} {\rm cm}^{-3}$ for all events.
The ambient plasma density $n_{i}$ is calculated using the thick
target model; it turns out to be $\sim 10^{10} {\rm cm}^{-3}$ for most events. Using these values for $n_{i}$, V,and the emission measure
we found typical background electron densities of  $ \sim 10^{10} {\rm cm}^{-3}$ (Table \ref{tab:fitparas}).   These numbers are around an order of
magnitude larger 
than those typically found for cold/tenuous flares \citep{Fleishman_2011}. The total thermal HXR energy thus calculated was found be  $\sim 10^{28}$ ergs
(Table \ref{tab:ennumbers}). 
%This
%is six orders of magnitude less than the total energy reported for big flares,
%indicating that the events could be microflares.


\begin{table*}
\caption{Energy chracteristics of Events
}
\begin{tabular}{cccccc}     % define the column alignment
                           % l: left, c: center, r: right
  \hline                   % horizontal line
 Date              & Energy in                      & Energy contained                 &  Energy carried away                        &Power                        &Power                                 \\
                   & thermal HXR emission                & in HXR electrons              &  by escaping electrons                    & in HXR electrons          & in escaping electrons                                   \\
                   &   (ergs)                             &     (ergs)                  &      (ergs)                                & (ergs/s)                           & (ergs/s)                                   \\
  \hline
Feb 28, 2004       & 1.06$\times10^{28}$            &  5.79$\times10^{26}$	            &    3.38$\times10^{25}$               & 1.69$\times10^{24}$             &9.07$\times10^{21}$                 \\
March 16,2004      & 1.19$\times10^{27}$            &  1.69$\times10^{24}$                  &    4.94$\times10^{24}$                & 1.18$\times10^{23}$            &8.78$\times10^{20}$  \\                                                                                                                  
June 26,2004       & 2.40$\times10^{27}$            &  9.94$\times10^{24}$                  &    1.05 $\times10^{23}$              &4.17$\times10^{23}$               &4.89$\times10^{21}$                     \\
June 27,2004       & 3.51$\times10^{27}$            &  1.18$\times10^{25}$                  &    9.55$\times10^{24}$               & 1.15$\times10^{23}$             &5.39$\times10^{21}$\\
Dec  25,2004       & 1.28$\times10^{27}$            &  2.16$\times10^{25}$                  &    7.32$\times10^{24}$                & 1.08$\times10^{23}$            &3.15$\times10^{21}$\\
Dec  25,2004       & 6.48$\times10^{26}$            &  1.02$\times10^{25}$                  &    6.05$\times10^{24}$                & 3.87$\times10^{23}$            &3.26$\times10^{21}$\\
\hline
\end{tabular}
\label{tab:ennumbers}
\end{table*}

% \section{Number flux, power and energy in accelerated electrons}

% {\color{red} YOU NEED TO CLEARLY SHOW HOW THE NUMBERS ARE CALCULATED}
 
%The number of electrons escaping to interplanetary space could be estimated
%by integrating over the average directional flux and energy (Lin 1971). The
%plot of number of HXR producing electrons and number of escaping electrons
%is given in Fig 3. We assumed electron spread cone of 30 following Lin(1974).
%Since all the events exhibited a rapid e-folding time with no time dependence on
%the decay profile it is reasonable to assume diffusion free propagation through
%interplanetary medium. Hence the number of escaped electrons was estimated
%by integrating over the solid angle spread and peak differential flux over the
%duration of the event. 

\section{Summary and Conclusions}
\subsection{Summary}
We have investigated the energy budgets involved in small electron acceleration events in the solar corona.
We envisage a scenario where these (small) acceleration events occur relatively high in the corona. Subsequently,
some of the accelerated electrons travel downwards and encounter the relatively dense chromospheric layers, resulting 
in the observed HXR emission via thick target bremsstrahlung. Some of the accelerated electrons escape outwards from the corona, 
travelling towards the Earth along the Parker spiral, and are detected in-situ at 1 AU as impulsive electron events by the ACE/EPAM detectors. 
We concentrate on events that give rise to faint, but observable HXR emission near the west limb, and are also associated with IP type III bursts
and impulsive electron spikes observed in-situ at 1 AU. These events are not associated with soft Xray flares $> C1$ or with CMEs. These selection 
criteria yield 6 events between  2004 and 2015. We use RHESSI data for HXR emission and analyze it using the standard RHESSI/OSPEX routines. 
For each event, we investigate the acceleration energy budget for the electrons that produce the observed HXR emission as well as the ones that
escape out of the corona to reach the Earth. 

\subsection{Conclusions}
In some respects (such as the lack of appreciable soft X-ray emission and the relatively low background thermal electron densities) 
the events we have shortlisted are similar to the cold and tenuous flares reported by \citep{Fleishman_2011,Fleishman_2016} .
We find that the electrons escaping from the acceleration site and detected at 1.2 AU have a power law law index ($\delta_{1}$) that is similar 
to the ones which produce the HXR emission ($\delta_{HXR}$). This suggests that the escaping electrons and the HXR producing electrons arise from the same population that was accelerated in the corona. The escaping electrons detected at 1 AU in the DE 4 channel take $\approx$ 14 minutes to travel from the Sun to the Earth, which suggests 
that scattering effects are not important (see Figure~\ref{fig:stacked}).

We compute the number, power and total energy in the nonthermal
electron populations producing the 1 AU electron spikes and the corresponding HXR emission using the prescriptions outlined 
in \S~3.1.1 and 3.2.1. 
Our main results are:
\begin{itemize}
\item
The ratio of the number of escaping electrons (that are detected in-situ at 1 AU by ACE/EPAM) to the number of
(downward precipitating) HXR producing electrons ranges from 6 \% to a number as large as 148 \% (Table \ref{tab:numbers}) . By comparison, \citet{krucker2007}, who analyzed much larger (typically M class) flares,
found a ratio of $\approx 0.2 \%$. 
On the other hand, based on simulations,\citet{Wang_2016} estimate that, in the quiet Sun, the ratio of electrons propagating outwards to form the solar 
wind superhalo 
to those that could potentially propagate downwards and produce HXR emission can be as high as 30 \%. 
%Our results strenghten this conclusion.
\item
The total energy in the electron spikes detected at 1 AU ranges from $\approx 10^{23}$ to $10^{25}$ erg, while that in the nonthermal electron
population producing the corresponding HXR events is $\approx 10^{24}$ -- $10^{26}$ erg (Table \ref{tab:ennumbers}). On the other hand, the energy contained in thermal HXR
emission is found to range from $10^{26}$-$10^{28}$ erg. 
\end{itemize}
Our analysis yields
insights pertaining to nonthermal electron acceleration in very small events, which can serve as a useful guide to understanding the feasibility 
of direct detections of nanoflares.Our results are also relevant to recent theories concerning particle acceleration in the turbulent solar corona. 
Astrophysical plasmas are expected to be generically turbulent; the solar corona and the solar wind are archetypal examples.
The presence of turbulence is expected to make the reconnection rate independent of microphysical (anomalous) resistivity and
engender fast reconnection (\citet{Lazarian_1999,Lazarian_2015} and references therein).
Alfvenic scattering centres in the vicinity of the turbulent reconnection sites are expected to aid in accelerating electrons 
\citep{Vlahos_2016}. Our results can constrain the parameters of such turbulence-aided acceleration scenarios.
For instance, the spectral slope of the high energy tail of the electrons detected in-situ at 1 AU
(which is fairly similar to that of the electrons producing HXR emission) can be used as a guide for determining the ratio of the
acceleration timescale to the escape timescale.

\section*{Acknowledgements}
%{\color{red} YOU NOW NEED TO PUT IN ALL THE REFERENCES PROPERLY}

TJ is thankful to IISER Pune for a PhD fellowship.
EPK acknowledges IISER Pune for hospitality during a visit in Nov 2016.EPK work was supported by a STFC consolidated grant ST/L000741/1.
We acknowledge a helpful and constructive review by the referee.  

%%%%%%%%%%%%%%%%%%%%%%%%%%%%%%%%%%%%%%%%%%%%%%%%%%



\bibliographystyle{mnras}
\bibliography{mnrasref} % if your bibtex file is called example.bib

%\begin{thebibliography}{99}

%\end{thebibliography}

%%%%%%%%%%%%%%%%%%%%%%%%%%%%%%%%%%%%%%%%%%%%%%%%%%

%%%%%%%%%%%%%%%%% APPENDICES %%%%%%%%%%%%%%%%%%%%%

%%%%%%%%%%%%%%%%%%%%%%%%%%%%%%%%%%%%%%%%%%%%%%%%%%

% Don't change these lines
\bsp	% typesetting comment
\label{lastpage}
\end{document}